\documentclass[9pt,twoside,twocolumn]{article}
\usepackage{extsizes}
\usepackage[super,sort&compress,comma]{natbib}
\usepackage[version=3]{mhchem}
\usepackage[left=1.5cm, right=1.5cm, top=1.785cm, bottom=2.0cm]{geometry}
\usepackage{balance}
\usepackage{widetext}
\usepackage{times,mathptmx}
\usepackage{sectsty}
\usepackage{graphicx}
\usepackage{lastpage}
\usepackage[format=plain,justification=raggedright,singlelinecheck=false,font={stretch=1.125,small,sf},labelfont=bf,labelsep=space]{caption}
\usepackage{float}
\usepackage{fancyhdr}
\usepackage{fnpos}
\usepackage[english]{babel}
\usepackage{array}
\usepackage{droidsans}
\usepackage{charter}
\usepackage[T1]{fontenc}
\usepackage[usenames,dvipsnames]{xcolor}
\usepackage{setspace}
\usepackage[compact]{titlesec}
\usepackage{amsmath,amssymb}
\definecolor{cream}{RGB}{222,217,201}
\begin{document}
%
\pagestyle{fancy}
\thispagestyle{plain}
\fancypagestyle{plain}{
\fancyhead[C]{}
\fancyhead[L]{\hspace{0cm}\vspace{1.5cm}}
\fancyhead[R]{\hspace{0cm}\vspace{1.7cm}}
\renewcommand{\headrulewidth}{0pt}}
\makeFNbottom
\makeatletter
\renewcommand\LARGE{\@setfontsize\LARGE{15pt}{17}}
\renewcommand\Large{\@setfontsize\Large{12pt}{14}}
\renewcommand\large{\@setfontsize\large{10pt}{12}}
\renewcommand\footnotesize{\@setfontsize\footnotesize{7pt}{10}}
\makeatother
\renewcommand{\thefootnote}{\fnsymbol{footnote}}
\renewcommand\footnoterule{\vspace*{1pt}%
\color{cream}\hrule width 3.5in height 0.4pt \color{black}\vspace*{5pt}}
\setcounter{secnumdepth}{5}
\makeatletter
\renewcommand\@biblabel[1]{#1}
\renewcommand\@makefntext[1]%
{\noindent\makebox[0pt][r]{\@thefnmark\,}#1}
\makeatother
\renewcommand{\figurename}{\small{Fig.}~}
\sectionfont{\sffamily\Large}
\subsectionfont{\normalsize}
\subsubsectionfont{\bf}
\setstretch{1.125}
\setlength{\skip\footins}{0.8cm}
\setlength{\footnotesep}{0.25cm}
\setlength{\jot}{10pt}
\titlespacing*{\section}{0pt}{4pt}{4pt}
\titlespacing*{\subsection}{0pt}{15pt}{1pt}
\fancyfoot{}
\fancyfoot[LO,RE]{\vspace{-7.1pt}}
\fancyfoot[CO]{\vspace{-7.1pt}\hspace{13.2cm}}
\fancyfoot[CE]{\vspace{-7.2pt}\hspace{-14.2cm}}
\fancyfoot[RO]{\footnotesize{\sffamily{1--\pageref{LastPage} ~\textbar  \hspace{2pt}\thepage}}}
\fancyfoot[LE]{\footnotesize{\sffamily{\thepage~\textbar\hspace{3.45cm} 1--\pageref{LastPage}}}}
\fancyhead{}
\renewcommand{\headrulewidth}{0pt}
\renewcommand{\footrulewidth}{0pt}
\setlength{\arrayrulewidth}{1pt}
\setlength{\columnsep}{6.5mm}
\setlength\bibsep{1pt}
\makeatletter
\newlength{\figrulesep}
\setlength{\figrulesep}{0.5\textfloatsep}
\newcommand{\topfigrule}{\vspace*{-1pt}%
\noindent{\color{cream}\rule[-\figrulesep]{\columnwidth}{1.5pt}}}
\newcommand{\botfigrule}{\vspace*{-2pt}%
\noindent{\color{cream}\rule[\figrulesep]{\columnwidth}{1.5pt}}}
\newcommand{\dblfigrule}{\vspace*{-1pt}%
\noindent{\color{cream}\rule[-\figrulesep]{\textwidth}{1.5pt}}}
\makeatother
%
\twocolumn[
\begin{@twocolumnfalse}\vspace{3cm}\sffamily
\begin{tabular}{m{4.5cm}p{13.5cm}}
&
\noindent\LARGE{\textbf{Hydrodynamic instabilities in shear flows of cohesive granular particles}}\\
\vspace{0.3cm}&\vspace{0.3cm}\\
&\noindent\large{Kuniyasu Saitoh,$^{\ast}$\textit{$^{a}$} Satoshi Takada,\textit{$^{b}$} and Hisao Hayakawa\textit{$^{b}$}}\\
&
\noindent\normalsize{
We extend the dynamic van der Waals model introduced by A. Onuki [\textit{Phys.\ Rev.\ Lett.\ } \textbf{94},\ 054501 (2005)]
to the description of cohesive granular flows under a plane shear to study their hydrodynamic instabilities.
Numerically solving the dynamic van der Waals model, we observe various heterogeneous structures of the density in steady states,
where the viscous heating is balanced with the energy dissipation caused by inelastic collisions.
Based on the linear stability analysis, we find that the spatial structures are determined
by the mean volume fraction, the applied shear rate, and the inelasticity,
where the instability is triggered if the system is \emph{thermodynamically unstable},\
i.e.\ the pressure, $p$, and the volume fraction, $\phi$, satisfy $\partial p/\partial\phi<0$.}\\
\end{tabular}
\end{@twocolumnfalse}\vspace{0.6cm}
]
\renewcommand*\rmdefault{bch}\normalfont\upshape
\rmfamily
\section*{}
\vspace{-1cm}
\footnotetext{\textit{$^{a}$~Faculty of Engineering Technology, MESA+, University of Twente, Drienerlolaan 5, 7522 NB, Enschede, The Netherlands. E-mail: k.saitoh@utwente.nl}}
\footnotetext{\textit{$^{b}$~Yukawa Institute for Theoretical Physics, Kyoto University, Sakyo-ku, Kyoto, 606-8502, Japan.}}
\section{Introduction}
\label{sec:intro}
Because flows of granular materials are ubiquitous in nature, a better understanding of their properties is crucially important in industry and science \cite{review1}.
In contrast to usual fluids, inelastic collisions between granular particles significantly influence the dynamics of granular flows \cite{review2,review3,nishant}.
As a result, the granular flow depends on both the volume fraction and externally applied force such as shear rate,
where many continuum models have been proposed to describe their anomalous rheology such as
kinetic theory of granular gases under shear \cite{kinetic_shear0,kinetic_shear1,kinetic_shear2,saitoh0},
constitutive relations for dense granular flows,\ i.e.\ the so-called $\mu$-$I$ rheology \cite{muI0,muI1,muI2,muI3,muI4,muI5,muI6,saitoh4},
revised non-local models for slow flows of granular materials \cite{nonl0,nonl1,nonl2},
and order-parameter descriptions for multiphase (fluid-solid coexistence) flows of granular particles
\cite{TDGLmodel0,TDGLmodel1,TDGLmodel2,TDGLmodel3,TDGLmodel4,TDGLmodel5,TDGLmodel6,TDGLmodel7,TDGLmodel8,TDGLmodel9,coexistence_granular0,coexistence_granular1}.
The interaction between grains is also an important factor in flows of granular materials,\ e.g.\
discontinuous shear thickening of frictional granular particles \cite{thickening0,thickening1,thickening2,thickening3},
strong shear resistance of dense cohesive granular particles \cite{CohesiveFlow,saitoh3},
jammed regime in the flow curve of dense cohesive granular materials \cite{CohesiveFlow1,CohesiveFlow3},
and instability of freely falling cohesive granular streams \cite{CohesiveFlow2}.

Among such a wide range of theoretical and numerical approaches for granular flows,
kinetic theory is one of the most successful methods in describing the hydrodynamics of dry granular particles
\cite{kinetic0,kinetic1,kinetic2,kinetic3,kinetic4,kinetic5,kinetic6,kinetic7,kinetic8,kinetic_friction0,kinetic_friction1}.
Though the underlying assumption seems to restrict its applicability (e.g.\ the contact duration should be zero),
kinetic theory gives quantitatively correct predictions of the transport coefficients for rigid granular particles
even for moderately dense systems \cite{kinetic_test1,kinetic_test}.
However, the basic assumption is violated once granular particles are aggregated \cite{aggregate1,aggregate2,aggregate3,aggregate4,aggregate5,aggregate6},
which is inavoidable for a collection of cohesive granular particles.
Note that the cohesive forces can have two different physical origins,\ i.e.\ \emph{van der Waals forces} between microscopic powders
or capillary forces between wet granular particles \cite{Castellanos,wetgranular}
\footnote[2]{In the following, we refer to the van der Waals interactions as the cohesive forces.}.
Recently, we studied the flows of cohesive granular particles under a plane shear using molecular dynamics simulations,
where we observed various spatial patterns caused by heterogeneous aggregates in steady states \cite{stakada0}.
Since such a heterogeneity cannot be explained by the conventional stability analysis of dry granular flows
\cite{linear0,linear1,linear2,linear3,linear4,linear5,linear6,linear7,linear8,linear9,shukla1,shukla2,shukla3,saitoh1,saitoh2},
it is a challenging task to explain the hydrodynamic instabilities in cohesive granular flows.

In this paper, we propose an extended dynamic van der Waals model originally proposed by A. Onuki \cite{waals1,waals2}
to describe hydrodynamic behaviors of a collection of cohesive granular particles.
Then, we study hydrodynamic instabilities in shear flows of cohesive granular particles with the aid of the dynamic van der Waals model.
First, we introduce a continuum model of cohesive granular particles in Sec.\ \ref{sec:model},
where we modify the dynamic van der Waals theory \cite{waals1,waals2} to include the energy dissipation caused by inelastic collisions between granular particles.
Then, we numerically solve the model under a plane shear in Sec.\ \ref{sec:numerics},
where we use the explicit MacCormack scheme \cite{mac} for numerical integrations and adopt the Lees-Edwards boundary condition \cite{lees,lees-FEM,lees-latticeB}.
In Sec.\ \ref{sec:stability}, we analyze the linear stability of homogeneous state to explain observed spatial structures in the presence of a shear rate and inelasticity.
Finally, we discuss and conclude our results in Secs.\ \ref{sec:discuss} and \ref{sec:conclude}, respectively.
In Appendices, we derive linearized hydrodynamics for the stability analysis (Appendix \ref{app:linearized})
and show our perturbative calculations of the eigenvalue problem (Appendix \ref{app:perturbation}).
%
\section{Model}
\label{sec:model}
In this section, we introduce a continuum model of cohesive granular materials,
where the \emph{dynamic van der Waals theory} for multiphase fluids \cite{waals1,waals2} is extended to include the dissipation of energy.
First, we show hydrodynamic equations of cohesive granular particles (Sec.\ \ref{sub:hydro}) and explain our model of constitutive relations (Sec.\ \ref{sub:const}).
Second, transport coefficients in the hydrodynamic equations are given by the kinetic theory of granular gases,
where a dissipation rate is also introduced to represent the effect of inelastic collisions (Sec.\ \ref{sub:trans}).
Third, we nondimensionalize the hydrodynamic equations and show their homogeneous solution (Sec.\ \ref{sub:nond}).
%
\subsection{Hydrodynamic equations}
\label{sub:hydro}
Let us introduce hydrodynamic fields as the mass density, $\rho=mn$, velocity field, $\tilde{u}_i$,
and \emph{granular temperature}
\footnote[3]{The granular temperature is defined as $T=m\langle\left(\tilde{\mathbf{v}}-\tilde{\mathbf{u}}\right)^2\rangle/d_\mathrm{m}n$
with the velocity of granular particle, $\tilde{\mathbf{v}}$, and the local velocity field, $\tilde{\mathbf{u}}$.}, $T$,
where $m$, $n$, and $i=\tilde{x},\tilde{y},\tilde{z}$ are the particle mass, the number density, and the coordinate, respectively
\footnote[4]{The variables with the tilde denote quantities having the physical dimension,
while those without the tilde, which will be used later, basically denote dimensionless quantities.}.
Then, the continuity equation, the equation of momentum conservation, and the equation of granular temperature in $d_\mathrm{m}$-dimension are given by
\begin{eqnarray}
\frac{\mathcal{D}\rho}{\mathcal{D}\tilde{t}} &=& -\rho\tilde{\nabla}_i\tilde{u}_i~, \label{eq:dim-he1}\\
\rho\frac{\mathcal{D}\tilde{u}_i}{\mathcal{D}\tilde{t}} &=& \tilde{\nabla}_j\tilde{\sigma}_{ij}~, \label{eq:dim-he2}\\
\frac{d_\mathrm{m}}{2}n\frac{\mathcal{D}T}{\mathcal{D}\tilde{t}} &=& \tilde{\sigma}_{ij}\tilde{\nabla}_i\tilde{u}_j
-\tilde{\nabla}_i\tilde{q}_i-\frac{d_\mathrm{m}}{2}nT\tilde{\zeta}~, \label{eq:dim-he3}
\end{eqnarray}
respectively, where we have used the Einstein convention for the subscripts ($i,j=\tilde{x},\tilde{y},\tilde{z}$).
On the left-hand-sides of Eqs.\ (\ref{eq:dim-he1})-(\ref{eq:dim-he3}), the material derivative is introduced as
$\mathcal{D}/\mathcal{D}\tilde{t}=\partial/\partial\tilde{t}+\tilde{u}_i\tilde{\nabla}_i$
with the time derivative, $\partial/\partial\tilde{t}$, and gradient, $\tilde{\nabla}_i$.
The last term on the right-hand-side of Eq.\ (\ref{eq:dim-he3}) represents the energy dissipation in the bulk caused by inelastic collisions,
where we have introduced a \emph{dissipation rate} as $\tilde{\zeta}$.
\subsection{Constitutive relations}
\label{sub:const}
Next, we discuss the constitutive relations for the stress tensor, $\tilde{\sigma}_{ij}$, and the heat flux, $\tilde{q}_i$.
The stress tensor is divided into the \emph{viscous} and \emph{reversible} parts as
\begin{equation}
\tilde{\sigma}_{ij}=\tilde{\tau}_{ij}-\tilde{\pi}_{ij}~,
\label{eq:dim-sigma}
\end{equation}
where the viscous part is defined as
\begin{equation}
\tilde{\tau}_{ij} = \tilde{\eta}\left(\tilde{\nabla}_i\tilde{u}_j+\tilde{\nabla}_j\tilde{u}_i\right)
+\delta_{ij}\left(\tilde{\xi}-\frac{2}{d_\mathrm{m}}\tilde{\eta}\right)\tilde{\nabla}_k\tilde{u}_k
\label{eq:dim-tau}
\end{equation}
($k=\tilde{x},\tilde{y},\tilde{z}$) with the shear viscosity, $\tilde{\eta}$, and bulk viscosity, $\tilde{\xi}$.
In the dynamic van der Waals theory \cite{waals1,waals2}, the reversible part can be written as
\begin{equation}
\tilde{\pi}_{ij} = (\tilde{p}+\tilde{p}_1)\delta_{ij}+M\tilde{\nabla}_in\tilde{\nabla}_jn~,
\label{eq:dim-pi}
\end{equation}
where the static pressure is given by the \emph{van der Waals equation of state},
\begin{equation}
\tilde{p} = \frac{nT}{1-v_0n}-\varepsilon v_0n^2~,
\label{eq:dim-p}
\end{equation}
with the particle volume, $v_0$, and well-depth of the attractive potential for cohesive granular particles, $\varepsilon$.
In Eq.\ (\ref{eq:dim-pi}), the diagonal part, $\tilde{p}_1$, and higher order gradient, $M\tilde{\nabla}_in\tilde{\nabla}_jn$,
with the coupling constant, $M$, represent the increase of energy due to the existence of interfaces between two different phases.
In this paper, we adopt the model used in Refs.\ \cite{waals1,waals2} for the diagonal part,\ i.e.\
\begin{equation}
\tilde{p}_1 = -\frac{M}{2}|\tilde{\nabla}n|^2-Mn\tilde{\nabla}^2n~,
\end{equation}
where the coupling constant is assumed to be proportional to the temperature as $M=2d^2v_0T$
with the particle diameter, $d$, measured by the range of square-well potential
\footnote[5]{The complete form of the diagonal part is given by
$\tilde{p}_1=\{(nM'-M)/2\}|\tilde{\nabla}_in|^2-nM\tilde{\nabla}_i^2n-nT(\tilde{\nabla}_in)\tilde{\nabla}_i(M/T)$ with $M'=\partial M/\partial n$,
where the surface tension is given by $\varsigma=\int_{-\infty}^\infty M\left(dn_{eq}/dr\right)^2dr$ with the equilibrium density profile, $n_{eq}(r)$.
If the coefficient depends only on the temperature, the diagonal part is reduced to the one used in this paper \cite{waals1,waals2}.}.
It should be noted that the coupling term can be derived from a microscopic model for thermodynamic interfaces \cite{rowlinson},
but we phenomenologically use this expression, because the microscopic derivation for cohesive granular particles, so far, does not exist.

The heat flux is given by
\begin{equation}
\tilde{q}_i = -\tilde{\kappa}\tilde{\nabla}_iT-\tilde{\mu}\tilde{\nabla}_in~,
\label{eq:dim-qi}
\end{equation}
where the first term on the right-hand-side represents Fourier's law with the thermal conductivity, $\tilde{\kappa}$.
The second term on the right-hand-side of Eq.\ (\ref{eq:dim-qi}), which does not exist in usual fluids, is derived from the kinetic theory of granular gases.
The physical origin of this term can be explained as follows:
Inelastic collisions in dense regions decrease the kinetic energy of granular particles so that the granular temperature tends to be lower than that in dilute regions
\cite{kinetic0,kinetic1,kinetic2,kinetic3,kinetic4,kinetic5,kinetic6,kinetic7,kinetic8,kinetic_shear0,kinetic_shear1,kinetic_shear2,kinetic_friction0,kinetic_friction1}.
%
\subsection{Transport coefficients and the dissipation rate}
\label{sub:trans}
Transport coefficients and the dissipation rate of moderately dense dry granular particles are well described by the kinetic theory
\cite{kinetic0,kinetic1,kinetic2,kinetic3,kinetic4,kinetic5,kinetic6,kinetic7,kinetic8,kinetic_friction0,kinetic_friction1}.
However, it is still a challenging task to derive those for cohesive granular particles,
where our attempt to develop a kinetic theory of cohesive granular gases is in progress \cite{stakada1}.
In this paper, we only study moderately dense systems, where the mean volume fraction of granular particles is much lower than $0.5$
(but is sufficiently dense to be regarded as a finite density system).
In addition, we assume that the granular particles are nearly elastic and are driven by a small shear rate to keep the low granular temperature.
We have already confirmed that the transport coefficients and the dissipation rate of cohesive granular particles
are well approximated by expanding the interaction range of a square-well potential with an inelastic repulsive hard-core,
at least, for nearly elastic dilute granular gases \cite{stakada1}.
Therefore, we use the transport coefficients and the dissipation rate derived from the kinetic theory of inelastic hard-core potentials,
where the diameter, $d$, represents the interaction range of the square-well potential.

From the kinetic theory of three-dimensional hard-core granular gases \cite{kinetic7},
the bulk viscosity, shear viscosity, and thermal conductivity are given by
\begin{eqnarray}
\tilde{\xi} &=& \frac{5f_\xi(\phi)}{16d^2}\sqrt{\frac{mT}{\pi}}~, \label{eq:dim-xi}\\
\tilde{\eta} &=& \frac{5f_\eta(\phi)}{16d^2}\sqrt{\frac{mT}{\pi}}~, \label{eq:dim-eta}\\
\tilde{\kappa} &=& \frac{75f_\kappa(\phi)}{64d^2}\sqrt{\frac{T}{\pi m}}~, \label{eq:dim-kappa}
\end{eqnarray}
respectively, where the explicit forms of dimensionless functions of the volume fraction, $\phi=v_0 n$,\
i.e.\ $f_\xi(\phi)$, $f_\eta(\phi)$, and $f_\kappa(\phi)$, are listed in Table \ref{tab:dfunc}.
Note that their dependences on the temperature, $\sqrt{T}$, are identical to those in usual hard-core fluids.
The transport coefficient for the density gradient in the heat flux, Eq.\ (\ref{eq:dim-qi}), is given by
\begin{equation}
\tilde{\mu} = \frac{75f_\mu(\phi)d}{64\sqrt{\pi m}}T^{3/2}
\label{eq:dim-mu}
\end{equation}
with the dimensionless function, $f_\mu(\phi)$, introduced in Table \ref{tab:dfunc}.

The dissipation rate is simply explained by \emph{Haff's law} \cite{kinetic0}:
The decrease of granular temperature by inelastic collisions is proportional to $(1-e^2)T$ with the restitution coefficient of granular particles, $e$.
The number of collisions per unit time is roughly estimated as $\chi(\phi)n\sqrt{T}$, where $\chi(\phi)$ is the radial distribution function at contact.
Then, the decrease of temperature per unit time is given by $\partial T/\partial\tilde{t}\sim-(1-e^2)nT^{3/2}\equiv-nT\tilde{\zeta}_\mathrm{H}$ if we assume $\chi(\phi)\sim1$.
More precise calculation by kinetic theory \cite{kinetic7} shows the existence of an additional term proportional to the velocity gradient
such that the total dissipation rate is found to be
\begin{equation}
\tilde{\zeta}=\tilde{\zeta}_\mathrm{H}+(1-e^2)f_\zeta(\phi)\tilde{\nabla}_i\tilde{u}_i~,
\label{eq:Haff}
\end{equation}
where the dimensionless function, $f_\zeta(\phi)$, is expressed in Table \ref{tab:dfunc}.
The first term on the right-hand-side corresponds to Haff's law, where its explicit form is given by
\begin{equation}
\tilde{\zeta}_\mathrm{H} = \frac{4d^2}{3}\left(1+\frac{3h_1(e)}{32}\right)(1-e^2)n\chi(\phi)\sqrt{\frac{\pi T}{m}}
\end{equation}
with the dimensionless coefficient, $h_1(e)$, as given in Table \ref{tab:dfunc}.
%
\begin{table}[h]\small
\caption{\
Dimensionless coefficients and dimensionless functions in the transport coefficients,
where $\chi(\phi)$ is the radial distribution function at contact.
Here, we have introduced a scaled pressure and derivative of the radial distribution function
as $p^\ast\equiv p/(\phi\theta)=1/(1-\phi)-\phi/\theta$ and $\chi_\phi\equiv\partial\chi/\partial\phi$, respectively.}
\label{tab:dfunc}
\begin{tabular*}{0.5\textwidth}{@{\extracolsep{\fill}}l}
\hline
$h_1(e)=\frac{32(1-e)(1-2e^2)}{81-17e+30e^2(1-e)}$ ~,\\
$h_2(e)=\frac{5}{24}(1-e^2)\left(1+\frac{3h_1}{32}\right)$~,\\
$h_3(e)=\left\{1-\frac{(1-e)^2}{4}\right\}\left(1-\frac{h_1}{64}\right)$~,\\
$h_4(e)=\frac{1+e}{3}\left\{1+\frac{33}{16}(1-e)+\frac{(19-3e)h_1}{1024}\right\}$~,\\
$h_5(e)=\frac{1+e}{3}\left[2e-1+\left\{\frac{1+e}{2}-\frac{5}{3(1+e)}\right\}h_1\right]$~,\\
$h_6(e)=\frac{1+e}{3}\left\{\frac{5(1-e)(9h_1^2+240h_1+52)}{4096}-\frac{15e^2(1-e)-498e+434}{1024}h_1+\frac{15e^2(1-e)-96e+128}{16}\right\}$~,\\
$\chi(\phi)=\left(1-\frac{\pi\phi}{12}\right)\left(1-\frac{\pi\phi}{6}\right)^{-3}$~,\\
$\nu(\phi)=\frac{\pi}{5}(1+e)\phi\chi(\phi)$~,\\
$f_\eta^\mathrm{k}(\phi)=(h_3-h_2)^{-1}\left\{\chi(\phi)^{-1}+e-\frac{1}{3}\right\}$~,\\
$f_\kappa^\mathrm{k}(\phi)=(h_4-4h_2)^{-1}\left\{\frac{(p^\ast+1)h_1+2}{3}\chi(\phi)^{-1}+h_5\right\}$~,\\
$f_\xi(\phi)=\frac{32-h_1}{9}\phi\nu(\phi)$~,\\
$f_\eta(\phi)=f_\eta^\mathrm{k}(\phi)\left(1+\frac{2}{3}\nu(\phi)\right)+\frac{3}{5}f_\xi(\phi)$~,\\
$f_\kappa(\phi)=f_\kappa^\mathrm{k}(\phi)\left(1+\nu(\phi)\right)+\frac{64+14h_1}{45}\phi\nu(\phi)$~,\\
$f_\mu(\phi)=\frac{1+\nu(\phi)}{5(h_4-3h_2)\chi(\phi)}\Big[\frac{1}{3}\frac{\partial(\phi p^\ast)}{\partial\phi}+\frac{5}{12}(1-e^2)\left(1+\frac{3h_1}{32}\right)
\frac{\partial(\phi\chi)}{\partial\phi}f_\kappa^\mathrm{k}(\phi)$ \\
$\hspace{2cm}-\frac{2\nu}{3}\left\{(1-e)e+\frac{4+3e-3e^2}{12}h_1\right\}\left(1+\frac{\phi\chi_\phi}{2\chi}\right)\Big]$~,\\
$f_\zeta(\phi)=\frac{1-p^\ast}{1+e}+\frac{5}{32h_6}\left(1+\frac{3h_1}{64}\right)\Big[(1-p^\ast)\left(e-\frac{2}{3}\right)h_1$ \\
$\hspace{2cm}+\left\{\frac{(1-e)(5e^2+4e-1)}{12}-\frac{(15e^2-3e-140)eh_1}{144}\right\}\nu(\phi)\Big]$~.\\
\hline
\end{tabular*}
\end{table}
%
\subsection{Nondimensionalization}
\label{sub:nond}
We introduce scaling units of the mass, length, energy, and time as the particle mass, $m$, the particle diameter, $d$,
the well-depth of the attractive potential for cohesive granular particles, $\varepsilon$, and a microscopic time scale,
$t_\mathrm{m}\equiv d(m/\varepsilon)^{1/2}$, respectively, so that the shear rate, $\dot{\gamma}$, is scaled as
\begin{equation}
s\equiv t_\mathrm{m}\dot{\gamma}~.
\end{equation}
Dimensionless hydrodynamic fields are introduced as the volume fraction, $\phi=v_0n$, dimensionless velocity field,
$u_i=(t_\mathrm{m}/d)\tilde{u}_i$, and dimensionless granular temperature, $\theta=T/\varepsilon$, respectively.
Then, the hydrodynamic equations (\ref{eq:dim-he1})-(\ref{eq:dim-he3}) are nondimensionalized as
\begin{eqnarray}
\frac{\mathcal{D}\phi}{\mathcal{D}t} &=& -\phi\nabla_iu_i~, \label{eq:he1}\\
\phi\frac{\mathcal{D}u_i}{\mathcal{D}t} &=& \nabla_j\sigma_{ij}~, \label{eq:he2}\\
\frac{d_\mathrm{m}}{2}\phi\frac{\mathcal{D}\theta}{\mathcal{D}t}
&=& \sigma_{ij}\nabla_iu_j - \nabla_i q_i - \frac{d_\mathrm{m}}{2}\phi\theta\zeta~, \label{eq:he3}
\end{eqnarray}
respectively, where we have used the Einstein convention for the dimensionless coordinates ($i,j=x,y,z$)
and have introduced the dimensionless material derivative as $\mathcal{D}/\mathcal{D}t\equiv\partial/\partial t+u_i\mathbf{\nabla}_i$
with $\partial/\partial t=t_\mathrm{m}\partial/\partial\tilde{t}$ and $\mathbf{\nabla}_i=d\tilde{\nabla}_i$.
In Table \ref{tab:dimensionless}, we summarize dimensionless forms of the stress tensor, the heat flux, the transport coefficients, and the dissipation rate.
%
\begin{table}[h]\small
\caption{\
Dimensionless forms of the stress tensor, $\sigma_{ij}=(v_0/\varepsilon)\tilde{\sigma}_{ij}$,
the heat flux, $q_i=(v_0t_\mathrm{m}/\varepsilon d)\tilde{q}_i$, the transport coefficients,\ i.e.\ $\xi=(v_0t_\mathrm{m}/md^2)\tilde{\xi}$,
$\eta=(v_0t_\mathrm{m}/md^2)\tilde{\eta}$, $\kappa=(t_\mathrm{m}d)\tilde{\kappa}$, and $\mu=(t_\mathrm{m}/\varepsilon d^2)\tilde{\mu}$,
and the dissipation rate, $\zeta=t_\mathrm{m}\tilde{\zeta}$, where the viscous stress, the reversible stress, the static pressure,
and the diagonal part of the reversible stress are nondimensionalized as $\tau_{ij}=(v_0/\varepsilon)\tilde{\tau}_{ij}$,
$\pi_{ij}=(v_0/\varepsilon)\tilde{\pi}_{ij}$, $p=(v_0/\varepsilon)\tilde{p}$, and $p_1=(v_0/\varepsilon)\tilde{p}_1$, respectively.}
\label{tab:dimensionless}
\begin{tabular*}{0.5\textwidth}{@{\extracolsep{\fill}}l}
\hline
$\sigma_{ij} = \tau_{ij}-\pi_{ij}$~,\\
$q_i = -\kappa\nabla_i\theta-\mu\nabla_i\phi$~,\\
$\tau_{ij} = \eta\left(\nabla_i u_j+\nabla_j u_i \right) + \delta_{ij}\left(\xi-2\eta/d_\mathrm{m}\right)\nabla_k u_k$~,\\
$\pi_{ij} = (p+p_1)\delta_{ij} + 2\theta\nabla_i\phi\nabla_j\phi$~,\\
$p = \phi\theta/(1-\phi)-\phi^2$~,\\
$p_1 = -\theta|\nabla\phi|^2-2\phi\theta\nabla^2\phi$~,\\
$\xi = (5/16\pi^{1/2})f_\xi(\phi)\sqrt{\theta}$~,\\
$\eta = (5/16\pi^{1/2})f_\eta(\phi)\sqrt{\theta}$~,\\
$\kappa = (75/64\pi^{1/2})f_\kappa(\phi)\sqrt{\theta}$~,\\
$\mu = (75/64\pi^{1/2})f_\mu(\phi)\theta^{3/2}$~,\\
$\zeta = \zeta_\mathrm{H}+(1-e^2)f_\zeta(\phi)\nabla_iu_i$~,\\
$\zeta_\mathrm{H} = (4\pi^{1/2}/3)(1+3h_1/32)(1-e^2)\phi\chi(\phi)\sqrt{\theta}$~.\\
\hline
\end{tabular*}
\end{table}

It is readily found that the dimensionless hydrodynamic equations (\ref{eq:he1})-(\ref{eq:he3}) have a homogeneous solution,
$\phi=\phi_0$, $\theta=\theta_0$, and $\mathbf{u}=\mathbf{u}_0\equiv(sy,0,0)$, corresponding to a uniform shear flow,
where $\phi_0$, $\theta_0$, and $\mathbf{u}_0$ are a homogeneous volume fraction, homogeneous temperature, and uniformly sheared velocity field, respectively.
From Eq.\ (\ref{eq:he3}), the homogeneous temperature is found to be
\begin{equation}
\theta_0 = \left\{\frac{15f_\eta(\phi_0)}{\pi d_\mathrm{m}(3h_1+32)\phi_0^2\chi(\phi_0)}\right\}\frac{s^2}{1-e^2}~.
\label{eq:theta0}
\end{equation}
Note that a finite value of the homogeneous temperature represents the balance between the \emph{viscous heating} and the \emph{dissipation of energy},
where the dimensionless shear rate and inelasticity are scaled as $s^2\sim1-e^2$
\footnote[6]{Note that the homogeneous temperature is an increasing function of time if there is no dissipation of energy,
where the increase of temperature is equal to the viscous heating,\ i.e.\ $\partial\theta_0(t)/\partial t=s^2\eta_0$.}.
%
\section{Numerical simulations}
\label{sec:numerics}
In this section, we numerically solve the dimensionless hydrodynamic equations (\ref{eq:he1})-(\ref{eq:he3}) under a plane shear.
We explain our numerical setup in Sec.\ \ref{sub:setup} and show our numerical results in Sec.\ \ref{sub:transient}.
\subsection{Setup}
\label{sub:setup}
We prepare a periodic $L\times L\times L$ cubic box with the dimensionless system size, $L/d=50$,
and divide it into $N=125000$ ($=50^3$) small cells with the identical volume, $d^3$.
Next, we randomly distribute the volume fraction, dimensionless temperature, and dimensionless velocity field in each cell around the homogeneous solution,\
i.e.\ $\phi_0$, $\theta_0$, and $\mathbf{u}_0=(sy,0,0)$, respectively, where the amplitudes of fluctuations are less than $10\%$ of the mean values.
Then, the \emph{explicit MacCormack scheme} \cite{mac} is used for numerical integrations of the dimensionless hydrodynamic equations (\ref{eq:he1})-(\ref{eq:he3}),
where the dimensionless time increment is fixed to be $\Delta t/t_\mathrm{m}=0.1$.

\begin{figure}[h]
\centering
\includegraphics[width=6cm]{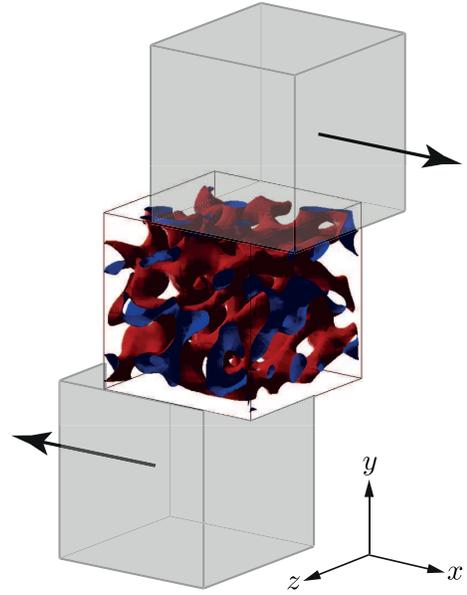}%
\caption{(Color online)
A sketch of our numerical setup.
The centered cube represents the bulk, where the isosurface corresponds to the volume fraction, $\phi_\mathrm{iso}=0.35$.
The volume fraction in the red (blue) side on the isosurface is lower (higher) than $\phi_\mathrm{iso}$.
The gray-shaded cubes are copies of the bulk,\ i.e.\ \emph{image-cells}, moving in the opposite directions
along the $x$-axis (indicated by the arrows) to apply a plane shear to the bulk.}
\label{fig:setup}
\end{figure}

To apply a plane shear to the system, we use the \emph{Lees-Edwards boundary condition}
which is originally proposed for molecular dynamics simulations \cite{lees} and is extended to
the finite-element method \cite{lees-FEM} and the lattice Boltzmann method \cite{lees-latticeB}.
Figure \ref{fig:setup} is a sketch of our numerical setup, where the centered cube and gray-shaded cubes represent the bulk and copies of the bulk (\emph{image-cells}), respectively.
In this figure, the isosurface in the bulk corresponds to the volume fraction, $\phi_\mathrm{iso}=0.35$,
where the volume fractions in the red and blue sides on the isosurface are lower and higher than $\phi_\mathrm{iso}$, respectively.
Then, we move the upper and lower image-cells in the opposite directions along the $x$-axis so that the system is sheared by the scaled shear rate, $s=t_\mathrm{m}\dot{\gamma}$.
Note that our method is different from the \emph{remesh procedure} \cite{remesh} which corresponds to the \emph{Sllod algorithm} for molecular dynamics simulations,
because the external shear is applied only at the boundaries and there is no external force in the bulk.
%
\subsection{Transient dynamics and steady states}
\label{sub:transient}
\begin{figure*}
\centering
\includegraphics[width=18cm]{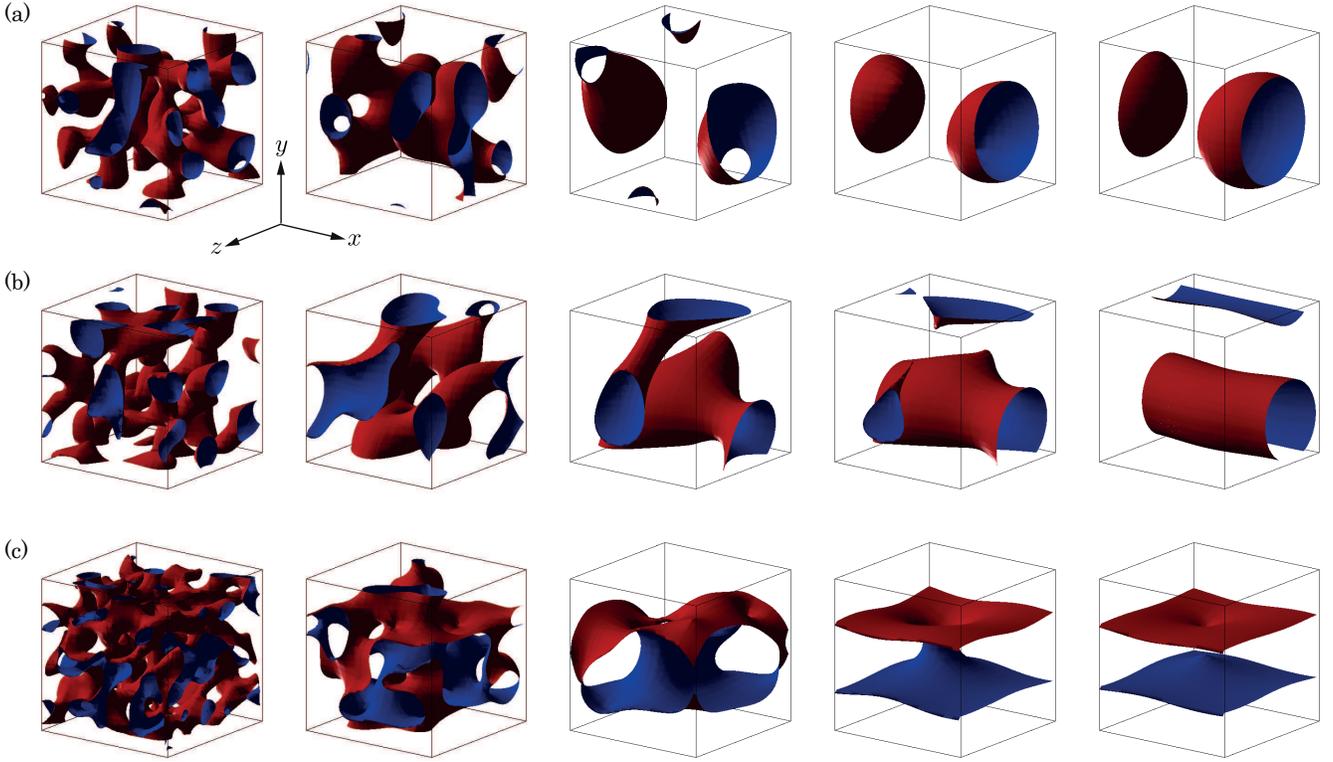}%
\caption{(Color online)
The time evolution of the isosurface for $\phi_\mathrm{iso}=0.35$, where the systems develop from \emph{left} to \emph{right}.
The volume fraction in the red (blue) side on the isosurface is lower (higher) than $\phi_\mathrm{iso}$.
The mean volume fractions are given by $\phi_0=$ (a) $0.8\phi_\mathrm{iso}$, (b) $0.9\phi_\mathrm{iso}$, and (c) $\phi_\mathrm{iso}$, respectively.
The dimensionless shear rate is fixed to $s=3\times10^{-4}$, while the inelasticities are, respectively, given by
$1-e^2=$ (a) $3.5\times10^{-7}$, (b) $3.0\times10^{-7}$, and (c) $2\times10^{-7}$.}
\label{fig:dynamics}
\end{figure*}

Depending on the mean volume fraction, $\phi_0$, dimensionless shear rate, $s$, and inelasticity, $1-e^2$,
the system exhibits various transient dynamics and different spatial structures in steady states.
Figure \ref{fig:dynamics} displays the time evolution of the isosurface, where the shear rate is fixed to be $s=3\times10^{-4}$.
In this figure, the mean volume fractions and the inelasticities are given by
$\phi_0=$ (a) $0.8\phi_\mathrm{iso}$, (b) $0.9\phi_\mathrm{iso}$, and (c) $\phi_\mathrm{iso}$,
and $1-e^2=$ (a) $3.5\times10^{-7}$, (b) $3.0\times10^{-7}$, and (c) $2\times10^{-7}$, respectively.
Initially, the isosurface has a random structure in space.
As time goes on, the density contrast starts to grow and the domains merge with each other to make a large cluster.
If the mean volume fraction is relatively low, the cluster is isolated in the bulk
so that we observe a spheroidal or a droplet like structure in the steady state (Fig.\ \ref{fig:dynamics}(a)).
On the other hand, if the mean volume fraction is relatively high, the cluster is elongated along the $x$-axis by the external shear
and we observe either a cylindrical structure (Fig.\ \ref{fig:dynamics}(b)) or a plate structure (Fig.\ \ref{fig:dynamics}(c)) in the steady state.
%
\pagebreak

\begin{figure}
\centering
\includegraphics[width=8cm]{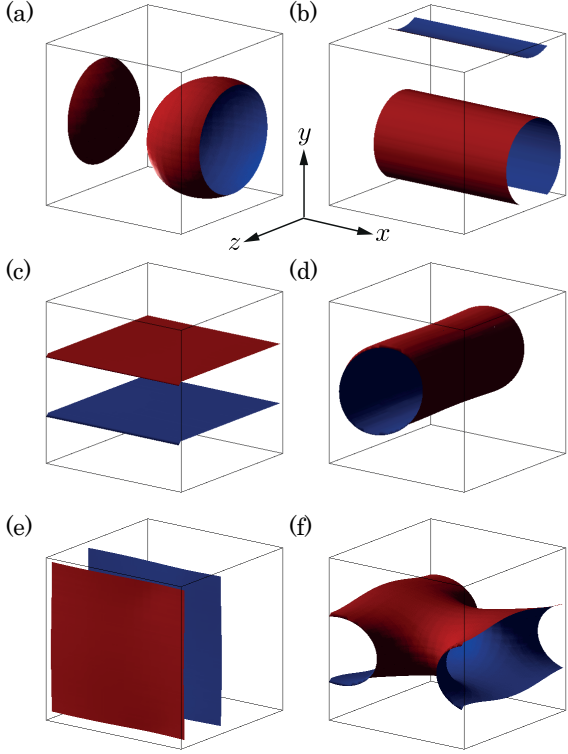}%
\caption{(Color online)
Typical structures of the isosurface in steady states,
where we classify them as (a) a \emph{droplet}, (b) a \emph{cylinder}, (c) a \emph{plate}, (d) a \emph{transverse-cylinder},
(e) a \emph{transverse-plate}, and (f) an \emph{irregular pattern}, respectively.}
\label{fig:pattern}
\end{figure}

We then classify spatial structures of the isosurface based on the dimensionless wave number, $(k_x,k_y,k_z)$, for the spatial undulation of the isosurface.
For example, $k_x=0$ if the isosurface is homogeneous along the $x$-axis, while $k_x=k_y=0$ if the isosurface is homogeneous along both the $x$- and $y$-axes, etc.
Clearly, the homogeneous state is characterized by $k_x=k_y=k_z=0$.
Figure \ref{fig:pattern} displays typical structures of the isosurface in steady states,
where we show (a) a \emph{droplet} ($k_x=k_y=k_z\neq0$), (b) a \emph{cylinder} ($k_x=0$, $k_y=k_z\neq0$), (c) a \emph{plate} ($k_x=k_z=0$, $k_y\neq0$),
(d) a \emph{transverse-cylinder} ($k_x=k_y\neq0$, $k_z=0$), and (e) a \emph{transverse-plate} ($k_x=k_y=0$, $k_z\neq0$) structure.
Here, we also introduce another case which does not belong to any of them as (f) an \emph{irregular pattern}.
%

\begin{figure}
\centering
\includegraphics[width=\columnwidth]{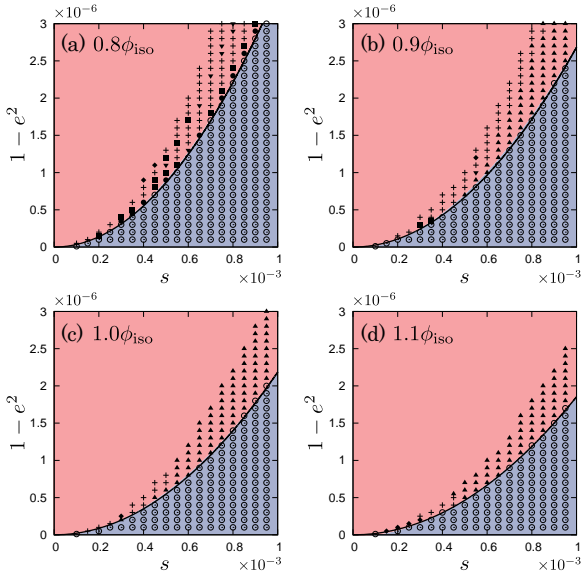}%
\caption{(Color online)
Phase diagrams of the spatial structures in the steady states plotted against the dimensionless shear rate, $s$, and inelasticity, $1-e^2$.
The mean volume fractions are fixed to $\phi_0=$ (a) $0.8\phi_\mathrm{iso}$, (b) $0.9\phi_\mathrm{iso}$, (c) $\phi_\mathrm{iso}$, and (d) $1.1\phi_\mathrm{iso}$, respectively.
The red (blue) region represents that the homogeneous state is unstable (stable).
Each spatial structure is classified as a \emph{homogeneous state} ($\odot$), a \emph{droplet} ($\bullet$),
a \emph{cylinder} ($\blacksquare$), a \emph{plate} ($\blacktriangle$), a \emph{transverse-cylinder} ($\blacktriangledown$),
a \emph{transverse-plate} ($\blacklozenge$), or an \emph{irregular pattern} ($+$).
The solid lines are the results of our linear stability analysis, Eq.\ (\ref{eq:neutral}).}
\label{fig:diag}
\end{figure}

Next, we map our numerical results onto phase diagrams of the dimensionless shear rate, $s$, and inelasticity, $1-e^2$.
Figure \ref{fig:diag} displays the phase diagrams for various mean volume fractions, $\phi_0$,
where both the spheroidal and cylindrical structures (\emph{droplet} and \emph{cylinder}) can be observed in relatively low volume fractions (Fig.\ \ref{fig:diag}(a)),
while the plate structures (\emph{plate} and \emph{transverse-plate}) appear in higher volume fractions (Figs.\ \ref{fig:diag}(b)-(d)).
In these figures, the initial homogeneous state is stable if the applied shear is large or the inelasticity is small,
where the borders between stable and unstable regions are well described by the solid lines obtained from our linear stability analysis in the next section (Sec.\ \ref{sec:stability}).
If the system is in the unstable region far from the solid line,\ i.e.\ in the highly nonlinear regime,
the structure in the steady state tends to be irregular and strongly depends on the initial condition,\ e.g.\ the pluses ($+$) in Figs.\ \ref{fig:diag}(a) and (b).
Note that we cannot simulate systems with highly inelastic situations,\ i.e.\ the parameter sets far above the solid lines in Fig.\ \ref{fig:diag} ($s^2\ll1-e^2$),
because the decrease of temperature is too fast to retain numerical stability.
%

\begin{figure}[h]
\centering
\includegraphics[width=6cm]{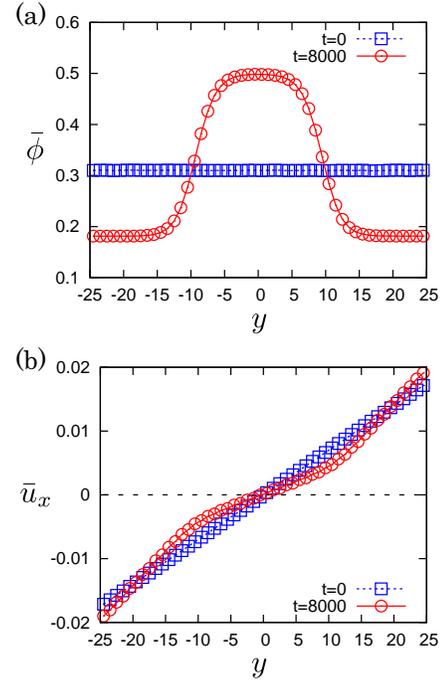}%
\caption{(Color online)
Profiles of (a) the volume fraction and (b) the dimensionless velocity field in the sheared direction plotted against the dimensionless coordinate, $y=\tilde{y}/d$,
where we average $\phi$ and $u_x$ over the $x$- and $z$-directions as Eqs.\ (\ref{eq:bar_phi_y}) and (\ref{eq:bar_ux_y}), respectively.
Here, the open squares and the open circles are the results in the initial state ($t=0$) and the steady state ($t=8000$), respectively,
where the mean volume fraction, dimensionless shear rate, and inelasticity are given by
$\phi_0=0.9\phi_\mathrm{iso}\simeq0.31$, $s=5\times10^{-4}$, and $1-e^2=7\times10^{-7}$, respectively.}
\label{fig:profiles}
\end{figure}

The system in the steady state is well sheared even though density contrast is observed in the bulk.
Figure \ref{fig:profiles} displays the profiles of volume fraction, $\bar{\phi}(y)$, and dimensionless velocity field in the sheared direction, $\bar{u}_x(y)$,
where we have averaged $\phi(x,y,z)$ and $u_x(x,y,z)$ over the $x$- and $z$-directions as
\begin{eqnarray}
\bar{\phi}(y) &=& \left(\frac{d}{L}\right)^2\iint\phi(x,y,z)dxdz~,\label{eq:bar_phi_y}\\
\bar{u}_x(y) &=& \left(\frac{d}{L}\right)^2\iint u_x(x,y,z)dxdz~,\label{eq:bar_ux_y}
\end{eqnarray}
respectively.
In this figure, the mean volume fraction, dimensionless shear rate, and inelasticity are given by
$\phi_0=0.9\phi_\mathrm{iso}\simeq0.31$, $s=5\times10^{-4}$, and $1-e^2=7\times10^{-7}$, respectively,
such that we observe a plate structure of the isosurface in the steady state (Fig.\ \ref{fig:pattern}(c)).
As shown in Fig.\ \ref{fig:profiles}(a), the initial homogeneous state, $\bar{\phi}(y)=\phi_0$ (the open squares), becomes \emph{unstable} by shear,
where the density in the steady state (the open circles) is divided into a dense region ($\bar{\phi}(y)\simeq0.5$)
and dilute regions ($\bar{\phi}(y)\lesssim0.2$) by the interfaces around the dimensionless coordinate, $y\simeq\pm10$.
As shown in Fig.\ \ref{fig:profiles}(b), the dimensionless velocity field well develops in the steady state (the open circles),
though it deviates from the initial linear velocity profile, $\bar{u}_x(y)=sy$ (the open squares).
%

In our molecular dynamics simulations of cohesive granular particles \cite{stakada0},
we observed the corresponding spatial structures to those displayed in Fig.\ \ref{fig:pattern},
where the phase diagrams are qualitatively identical to those in Fig.\ \ref{fig:diag}.
Moreover, the profiles of volume fraction and dimensionless velocity field are similar to those in Fig.\ \ref{fig:profiles}.
Though the range of scaled shear rate ($10^{-4}\le s\le10^{-3}$) is much smaller than
that studied in our molecular dynamics simulations ($10^{-4}\le s_\mathrm{MD}\le1$) \cite{stakada0},
our continuum model well captures the dynamics of cohesive granular particles under a plane shear.
%
\section{Linear stability analysis}
\label{sec:stability}
In this section, we analyze the linear stability of the homogeneous state to explain
the dependence of observed spatial structures on the \emph{control parameters},\ i.e.\ $\phi_0$, $s$, and $1-e^2$.

First, we add small fluctuations, $\hat{\phi}$, $\hat{\theta}$, and $\hat{\mathbf{u}}=(\hat{u}_x,\hat{u}_y,\hat{u}_z)$, to the homogeneous state
as $\phi=\phi_0+\hat{\phi}$, $\theta=\theta_0+\hat{\theta}$, and $\mathbf{u}=\mathbf{u}_0+\hat{\mathbf{u}}$, respectively.
Then, we linearize the dimensionless hydrodynamic equations (\ref{eq:he1})-(\ref{eq:he3}) against the small fluctuations.
For example, the left-hand-side of the continuity equation (\ref{eq:he1}) is linearized as
\begin{equation}
\frac{\partial\phi}{\partial t}+\mathbf{u}\cdot\nabla\phi \simeq \frac{\partial\hat{\phi}}{\partial t} + sy\nabla_x\hat{\phi}~.
\label{eq:example_lin_he1}
\end{equation}
Here, the second term on the right-hand-side of Eq.\ (\ref{eq:example_lin_he1}) is the result of
$\mathbf{u}_0\cdot\nabla\hat{\phi}=sy\nabla_x\hat{\phi}$ which explicitly depends on the coordinate, $y$.
With the aid of Eq.\ (\ref{eq:example_lin_he1}), we linearize Eqs.\ (\ref{eq:he1})-(\ref{eq:he3}) as
\begin{eqnarray}
\frac{\partial\hat{\phi}}{\partial t} + sy\nabla_x\hat{\phi} &=& -\phi_0\left(\nabla_x\hat{u}_x+\nabla_y\hat{u}_y+\nabla_z\hat{u}_z\right)~,
\label{eq:lin_he1}\\
\frac{\partial\hat{u}_x}{\partial t} + sy\nabla_x\hat{u}_x
&=& \left\{s\bar{\eta}_\phi\nabla_y+\left(2\theta_0\nabla^2-\bar{p}_\phi\right)\nabla_x\right\}\hat{\phi}\nonumber\\
&+& \left(s\bar{\eta}_\theta\nabla_y-\bar{p}_\theta\nabla_x\right)\hat{\theta}
+\left(\bar{\upsilon}_0\nabla_x^2+\bar{\eta}_0\nabla^2\right)\hat{u}_x\nonumber\\
&+& \left(\bar{\upsilon}_0\nabla_x\nabla_y-s\right)\hat{u}_y+ \bar{\upsilon}_0\nabla_z\nabla_x\hat{u}_z~,
\label{eq:lin_he2x}\\
\frac{\partial\hat{u}_y}{\partial t} + sy\nabla_x\hat{u}_y
&=& \left\{s\bar{\eta}_\phi\nabla_x+\left(2\theta_0\nabla^2-\bar{p}_\phi\right)\nabla_y\right\}\hat{\phi}\nonumber\\
&+& \left(s\bar{\eta}_\theta\nabla_x-\bar{p}_\theta\nabla_y\right)\hat{\theta}
+ \bar{\upsilon}_0\nabla_x\nabla_y\hat{u}_x\nonumber\\
&+& \left(\bar{\upsilon}_0\nabla_y^2+\bar{\eta}_0\nabla^2\right)\hat{u}_y+ \bar{\upsilon}_0\nabla_y\nabla_z\hat{u}_z~,
\label{eq:lin_he2y}\\
\frac{\partial\hat{u}_z}{\partial t} + sy\nabla_x\hat{u}_z
&=& \left(2\theta_0\nabla^2-\bar{p}_\phi\right)\nabla_z\hat{\phi}-\bar{p}_\theta\nabla_z\hat{\theta} + \bar{\upsilon}_0\nabla_z\nabla_x\hat{u}_x\nonumber\\
&+& \bar{\upsilon}_0\nabla_y\nabla_z\hat{u}_y + \left(\bar{\upsilon}_0\nabla_z^2+\bar{\eta}_0\nabla^2\right)\hat{u}_z~,
\label{eq:lin_he2z}\\
\frac{\partial\hat{\theta}}{\partial t} + sy\nabla_x\hat{\theta}
&=& \left(\bar{\mu}_0\nabla^2+\bar{\omega}_\phi\right)\hat{\phi} + \left(\bar{\kappa}_0\nabla^2+\bar{\omega}_\theta\right)\hat{\theta}\nonumber\\
&+& \left(bs\nabla_y-a\nabla_x\right)\hat{u}_x+\left(bs\nabla_x-a\nabla_y\right)\hat{u}_y- a\nabla_z\hat{u}_z~,\nonumber\\
\label{eq:lin_he3}
\end{eqnarray}
where the coefficients, $\bar{p}_\phi$, $\bar{p}_\theta$, $\bar{\eta}_0$, $\bar{\eta}_\phi$, $\bar{\eta}_\theta$,
$\bar{\omega}_\phi$, $\bar{\omega}_\theta$, $\bar{\upsilon}_0$, $\bar{\kappa}_0$, $\bar{\mu}_0$,
$a$, and $b$, are listed in Table \ref{tab:coefficients}.
%
\begin{table}[h]\small
\caption{\
Coefficients in the linearized hydrodynamic equations (\ref{eq:lin_he1})-(\ref{eq:lin_he3}),
where $\eta_0$, $\xi_0$, and $\kappa_0$ are $\eta$, $\xi$, and $\kappa$, in the homogeneous state, respectively.
The subscripts ($\phi$ and $\theta$) represent their derivatives in the homogeneous state,\ i.e.\ $p_\phi=\partial p/\partial\phi$,
$p_\theta=\partial p/\partial\theta$, $\eta_\theta=\partial\eta/\partial\theta$, $\eta_\theta=\partial\eta/\partial\theta$,
$\omega_\phi=\partial\omega/\partial\phi$, and $\omega_\theta=\partial\omega/\partial\theta$,
where $\omega\equiv s^2\eta-d_\mathrm{m}\phi\theta\zeta_\mathrm{H}/2$.}
\label{tab:coefficients}
\begin{tabular*}{0.5\textwidth}{@{\extracolsep{\fill}}ll}
\hline
$a_0=2p_\theta\theta_0/(d_\mathrm{m}\phi_0)$, & $\upsilon_0=(1-2/d_\mathrm{m})\eta_0+\xi_0$, \\
$\bar{\eta}_0=\eta_0/\phi_0$, & $\bar{\upsilon}_0=\upsilon_0/\phi_0$, \\
$\bar{\kappa}_0=2\kappa_0/(d_\mathrm{m}\phi_0)$, & $\bar{\mu}_0=2\mu_0/(d_\mathrm{m}\phi_0)$, \\
$\bar{p}_\phi=p_\phi/\phi_0$, & $\bar{p}_\theta=p_\theta/\phi_0$, \\
$\bar{\eta}_\phi=\eta_\phi/\phi_0$, & $\bar{\eta}_\theta=\eta_\theta/\phi_0$, \\
$\bar{\omega}_\phi=2\omega_\phi/(d_\mathrm{m}\phi_0)$, & $\bar{\omega}_\theta=2\omega_\theta/(d_\mathrm{m}\phi_0)$, \\
$b=4\eta_0/(d_\mathrm{m}\phi_0)$, & $a=a_0+(1-e^2)\theta_0f_\zeta(\phi_0)$. \\
\hline
\end{tabular*}
\end{table}

Second, we introduce the Fourier transforms as
\begin{eqnarray}
\hat{\phi} &=& \int\phi_\mathbf{k}(t)e^{i\mathbf{k}\cdot\mathbf{r}}d\mathbf{k}~,\label{eq:Fourier_phi}\\
\hat{\theta} &=& \int\theta_\mathbf{k}(t)e^{i\mathbf{k}\cdot\mathbf{r}}d\mathbf{k}~,\label{eq:Fourier_theta}\\
\hat{u}_j &=& i\int u_{j\mathbf{k}}(t)e^{i\mathbf{k}\cdot\mathbf{r}}d\mathbf{k}~,\label{eq:Fourier_u}
\end{eqnarray}
where $i$ and $\mathbf{k}=(k_x,k_y,k_z)$ are the imaginary unit and dimensionless wave number vector
(such that the wave number vector is given by $\tilde{\mathbf{k}}=\mathbf{k}/d$), respectively.
The Fourier transform of $sy\nabla_x\hat{\phi}$ in Eq.\ (\ref{eq:example_lin_he1}) is given by (see Appendix \ref{app:sub:Fourier} and Ref.\ \cite{onuki})
\begin{equation}
sy\nabla_x\hat{\phi} = -\int sk_x\frac{\partial\phi_\mathbf{k}}{\partial k_y}e^{i\mathbf{k}\cdot\mathbf{r}}d\mathbf{k}~.
\end{equation}
Thus, the Fourier transforms of the linearized hydrodynamic equations (\ref{eq:lin_he1})-(\ref{eq:lin_he3}) are written as
\begin{equation}
\left(\frac{\partial}{\partial t}-sk_x\frac{\partial}{\partial k_y}\right)\mathbf{\varphi}_\mathbf{k} = \mathcal{L}\mathbf{\varphi}_\mathbf{k}~,
\label{eq:lin_he}
\end{equation}
where $\mathbf{\varphi}_\mathbf{k}=\left(\phi_\mathbf{k},\theta_\mathbf{k},u_{x\mathbf{k}},u_{y\mathbf{k}},u_{z\mathbf{k}}\right)^\mathrm{T}$
is a transverse vector of the Fourier coefficients and $\mathcal{L}$ is a time-independent $5\times5$ matrix defined as Eq.\ (\ref{eq:L}) in Appendix \ref{app:sub:Fourier}.

Third, we introduce a \emph{growth rate} of the Fourier coefficients as $\hat{\mathbf{\varphi}}_\mathbf{k}(t)\propto e^{\lambda t}$
so that the linearized hydrodynamic equation (\ref{eq:lin_he}) is reduced to an eigenvalue problem,
\begin{equation}
\left(\mathcal{L}+sk_x\frac{\partial}{\partial k_y}\right)\mathbf{\varphi}_\mathbf{k} = \lambda\mathbf{\varphi}_\mathbf{k}~.
\label{eq:eigenvalue_problem}
\end{equation}
In Appendix \ref{app:perturbation}, we perturbatively solve the eigenvalue problem (\ref{eq:eigenvalue_problem})
by expanding the eigenvalues, eigenvectors, and matrix into the powers of the wave number, $k=|\mathbf{k}|$.
In our perturbative calculations, the shear rate and inelasticity are scaled as $s\sim O(k^2)$ and $1-e^2\sim O(k^4)$, respectively,
so that the homogeneous temperature, $\theta_0\sim s^2/(1-e^2)$, remains as finite.
Then, we find that the eigenvalue for the most unstable mode is given by $\lambda=\lambda^{(3)}$ with
\begin{equation}
\lambda^{(3)} \simeq -\frac{2\kappa_0p_\phi}{d_\mathrm{m}\phi_0f^2}k^2
\label{eq:most_unstable}
\end{equation}
(see Eq.\ (\ref{app:eq:isotropic_eigenvalue}) in Appendix \ref{app:sub:unstable_mode}),
where we have truncated the expansion of $\lambda^{(3)}$ at $k^2$
and have introduced a coefficient, $f=\sqrt{a_0\bar{p}_\phi+\phi_0\bar{p}_\theta}$.
Therefore, the eigenvalue is positive if
\begin{equation}
p_\phi=\frac{\partial p}{\partial\phi}<0~,
\end{equation}
i.e.\ the hydrodynamic instability is triggered if the system is \emph{thermodynamically unstable}.
Note that the other factor in Eq.\ (\ref{eq:most_unstable}) is negative, $-2\kappa_0k^2/d_\mathrm{m}\phi_0f^2<0$.
The \emph{neutral curve},\ i.e.\ $p_\phi=0$, is given by the van der Waals equation of state, Eq.\ (\ref{eq:dim-p}),
and the homogeneous granular temperature, Eq.\ (\ref{eq:theta0}),
where the dimensionless \emph{critical shear rate} for the neutral stability is found to be
\begin{equation}
s_\mathrm{cr} = \sqrt{\frac{2\pi d_\mathrm{m}\phi_0^3(1-\phi_0)^2\chi(\phi_0)\left\{3h_1(e)+32\right\}(1-e^2)}{15f_\eta(\phi_0)}}~.
\label{eq:neutral}
\end{equation}
The solid lines in the phase diagrams (Fig.\ \ref{fig:diag}) are given by Eq.\ (\ref{eq:neutral}) which well describe the results of numerical simulations.
Note that there is no fitting parameter in Eq.\ (\ref{eq:neutral}).

\begin{figure}[h]
\centering
\includegraphics[width=\columnwidth]{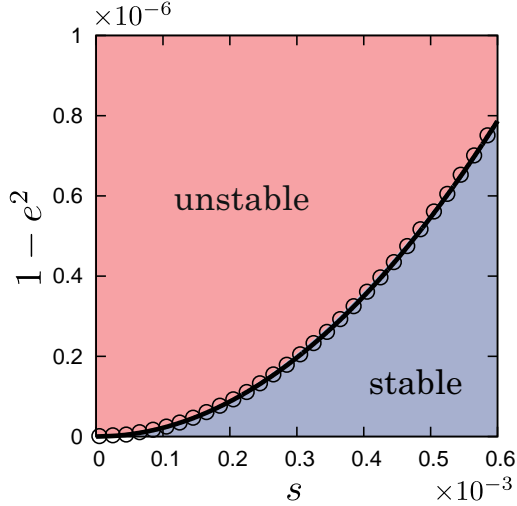}%
\caption{(Color online)
A stability diagram plotted against $s$ and $1-e^2$, where the mean volume fraction is fixed to $\phi_0=\phi_\mathrm{iso}$.
The unstable (red) and stable (blue) regions are divided by the neutral curve, Eq.\ (\ref{eq:neutral}) (the solid line),
where we obtain a good agreement with the numerical results (the open circles).}
\label{fig:neut_zerokx_curve}
\end{figure}

\begin{figure}[h]
\centering
\includegraphics[width=\columnwidth]{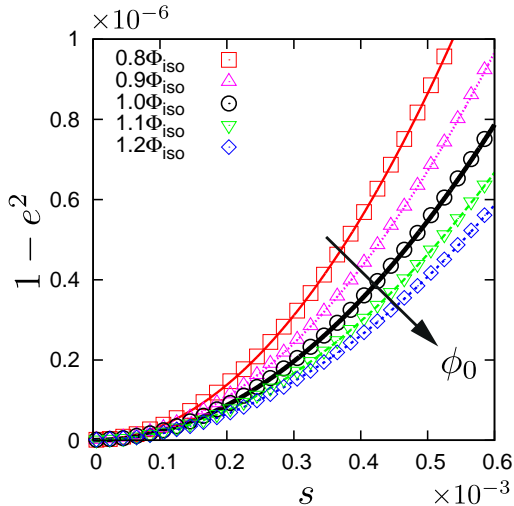}%
\caption{(Color online)
The neutral curves with different mean volume fractions, $\phi_0$, where all the numerical results (the open symbols) are well described by Eq.\ (\ref{eq:neutral}) (the lines).
Here, $\phi_0$ increases from $0.8\phi_\mathrm{iso}$ to $1.2\phi_\mathrm{iso}$ as listed in the legend and indicated by the arrow.}
\label{fig:neut_zerokx_curve_Theory-LAPACK}
\end{figure}

Our perturbative calculation also agrees with the numerical solution of the eigenvalue problem, Eq.\ (\ref{eq:eigenvalue_problem}).
Figure \ref{fig:neut_zerokx_curve} is a stability diagram plotted against the shear rate, $s$, and inelasticity, $1-e^2$,
where the solid line is the neutral curve, Eq.\ (\ref{eq:neutral}), and the open circles are numerical results of the critical shear rate.
Here, the LAPACK subroutines \cite{LAPACK} are used to numerically solve the eigenvalue problem, Eq.\ (\ref{eq:eigenvalue_problem}),
where we confirm a good agreement between our perturbative calculation and the numerical result.
As shown in Fig.\ \ref{fig:neut_zerokx_curve_Theory-LAPACK},
we also confirm that Eq.\ (\ref{eq:neutral}) well describes numerical results with different mean volume fractions,
where the unstable region increases with the increase of $\phi_0$.

It should be noted that the second term on the left-hand-side of the linearized hydrodynamic equation (\ref{eq:lin_he}),\
i.e.\ $-sk_x\partial\mathbf{\varphi}_\mathbf{k}/\partial k_y$, can be eliminated
by introducing the time-dependent wave number vector as $\mathbf{k}(t)=(k_x,k_y-stk_x,k_z)$,\ i.e.\ the \emph{Kelvin mode}.
In this case, however, we cannot use an ordinary procedure for the linear stability analysis,
where the $5\times5$ matrix becomes time-dependent, $\mathcal{L}(t)$, so that the eigenvalues also depend on time
and the eigenvectors have to be constructed of Green's function as $\varphi_\mathbf{k}(t)=\int G(\mathbf{k},\mathbf{k}',t)\varphi_{\mathbf{k}'}(0)d\mathbf{k}'$.
For the details of this method for dry granular flows, see our previous work in Ref.\ \cite{saitoh1}.
%
%
\section{Discussion}
\label{sec:discuss}
In this paper, we have studied hydrodynamic instabilities in a continuum model of cohesive granular particles under a plane shear.
The dynamic van der Waals theory for multiphase fluids \cite{waals1,waals2} has been extended to include the energy dissipation caused by inelastic collisions,
where the transport coefficients and dissipation rate derived from the kinetic theory of three-dimensional inelastic hard-core potential \cite{kinetic7} were used.

We have numerically solved the hydrodynamic equations for various values of the control parameters,\ i.e.\ $\phi_0$, $s$, and $1-e^2$,
where the explicit MacCormack scheme \cite{mac} was adopted for numerical integrations.
To apply a plane shear to the system, the Lees-Edwards boundary condition \cite{lees,lees-FEM,lees-latticeB} was used,
where there was no bulk shear in contrast to the remesh procedure \cite{remesh}.
Then, we observed heterogeneous structures of the density field in steady states,
where a \emph{spheroidal} or \emph{cylindrical} structure appeared if the mean volume fraction is relatively small,
while \emph{plate} structures appeared in the systems with higher volume fractions.
Note that such regular structures can be observed in the vicinity of the neutral curve, where we observed various \emph{irregular patterns} in highly nonlinear regimes.
All the spatial structures observed in our molecular dynamics simulations \cite{stakada0} have been reproduced by the hydrodynamic equations,
where the phase diagrams (Fig.\ \ref{fig:diag}) and the profiles (Fig.\ \ref{fig:profiles}) are qualitatively similar to those obtained in our previous study.

To explain the dependence of the spatial structures on the control parameters, we have analyzed the linear stability of the homogeneous state,
where we perturbatively solved the eigenvalue problem for the growth rate of small fluctuations.
From our linear stability analysis, we have found that the hydrodynamic instability is triggered if $p_\phi<0$,\ i.e.\ if the system is thermodynamically unstable.
Then, the boundaries between stable and unstable regions in the phase diagrams (Fig.\ \ref{fig:diag}) were well described by the neutral curve, Eq.\ (\ref{eq:neutral}),
where we also obtained a good agreement between Eq.\ (\ref{eq:neutral}) and the numerical result of the eigenvalue problem, Eq.\ (\ref{eq:eigenvalue_problem})
(Figs.\ \ref{fig:neut_zerokx_curve} and \ref{fig:neut_zerokx_curve_Theory-LAPACK}).

Though the neutral curve, Eq.\ (\ref{eq:neutral}), is given by the stability criterion, Eq.\ (\ref{eq:most_unstable}),
the eigenvalue, $\lambda^{(3)}\sim k^2$, is \emph{isotropic} in the Fourier space.
In other words, the isotropic eigenvalue cannot distinguish the observed spatial structures.
On the other hand, we also find the \emph{anisotropic} eigenvalue, $\lambda^{(4)}=se_xe_y-\bar{\eta}_0k^2$
(see Eq.\ (\ref{app:eq:anisotropic_eigenvalue}) in Appendix \ref{app:sub:unstable_mode}),
where its stability criterion, Eq.\ (\ref{app:eq:anisotropic_unstable_condition}),
corresponds to the \emph{shear-induced instability} for usual (dry) granular shear flows \cite{saitoh1}.
Therefore, the thermodynamic instability, $p_\phi<0$, and the shear-induced instability, Eq.\ (\ref{app:eq:anisotropic_unstable_condition}),
compete with each other, where Eq.\ (\ref{app:eq:anisotropic_unstable_condition}) also depends on the system size, $L$, through the wave numbers.
We find that the isotropic eigenvalue is always larger than the anisotropic one,\ i.e.\ $\lambda^{(3)}>\lambda^{(4)}$,
because our system size, $L=50d$, is too small to observe the shear-induced instability for the range of control parameters studied in this paper.
In future, further systematic studies of the \emph{pattern selection} for larger systems will be needed
as well as the \emph{weakly nonlinear analysis} for the amplitude equation \cite{shukla1,shukla2,shukla3,saitoh1}.
It should be noted that the temperature increases with time if there is no dissipation of energy.
Thus, the homogeneous solution is always linearly stable in the absence of inelastic collisions \cite{LutskoDufty}.
In our model, however, the mean temperature converges to a finite value in the steady state because the viscous heating is canceled by the energy dissipation.
Therefore, the hydrodynamic instability presented in this paper is one of consequences of the dissipative nature of granular materials.
We also stress that the thermodynamic instability, $p_\phi<0$, can be achieved only if the interaction between the particles is attractive.
In addition, the stability analyses of dry granular shear flows show that the hydrodynamic instability is induced only by the \emph{layering mode} ($k_x=0$),
while the non-layering mode ($k_x\neq0$) is always linearly stable \cite{linear5,linear6,linear7,linear8,shukla1,shukla2,shukla3,saitoh1}.
Therefore, spatial undulations in the sheared direction ($x$-axis),\ e.g.\ droplets (Fig.\ \ref{fig:pattern}(a)),
transverse-cylinders (Fig.\ \ref{fig:pattern}(d)), and irregular patterns (Fig.\ \ref{fig:pattern}(f)), do not exist in dry granular systems.
Thus, our results are also specific to cohesive granular materials.

Because we studied moderately dense systems with the mean volume fractions around $\phi_\mathrm{iso}=0.35$,
we have used the transport coefficients and dissipation rate derived from the kinetic theory of inelastic hard-core potentials \cite{kinetic7}.
This assumption may be validated if the externally applied shear rate is so small that the granular temperature stays in low values,
where the macroscopic properties of cohesive granular particles can be approximated by expanding the interaction range of a square-well potential \cite{stakada1}.
However, the microscopic determinations of realistic transport coefficients, the dissipation rate,
and the coupling constant, $M$, of cohesive granular materials are important,
where our attempt to develop a kinetic theory of cohesive granular gases is in progress \cite{stakada1}.

Moreover, the effects of gravity and microscopic frictions between the particles should be examined for practical applications,
and the influence of the boundary condition,\ e.g.\ the study with remesh procedure \cite{remesh} or physical boundary conditions, is also important.
%
%
\section{Conclusion}
\label{sec:conclude}
In conclusion, the extended dynamic van der Waals model can describe cohesive granular flows under a plane shear,
where the hydrodynamic instabilities are well characterized by the neutral curve obtained from the linearized hydrodynamics.
The various spatial structures observed in simulations appear in the unstable region,
where the hydrodynamic instabilities are triggered if the system is thermodynamically unstable,\ i.e.\ $p_\phi<0$.

\section*{Acknowledgements}
We thank M. Alam, K. Takae, H. Ebata, S. Nagahiro, and D. Vescovi for fruitful discussions.
We are grateful to N. Rivas for his critical reading and helpful comments on this manuscript.
K.S.\ wishes to express his gratitude to the Yukawa Institute for Theoretical Physics (YITP) for the support to his stay and its warm hospitality.
Part of this work was performed during the YITP workshops, ``\emph{Physics of Glassy and Granular Material}" (Grant No.\ YITP-W-13-04)
and ``\emph{Physics of Granular Flow}" (Grant No.\ YITP-T-13-03).
Numerical computation in this work was partially carried out at the Yukawa Institute Computer Facility.
This work was financially supported by the NWO-STW VICI Grant No.\ 10828 and JSPS KAKENHI Grant No.\ 25287098.


%
\appendix
\section{Linearized hydrodynamics}
\label{app:linearized}
%
%
In this Appendix, we linearize the dimensionless hydrodynamic equations (\ref{eq:he1})-(\ref{eq:he3})
around the homogeneous solution, $\phi=\phi_0$, $\theta=\theta_0$, and $\mathbf{u}=\mathbf{u}_0\equiv(sy,0,0)$,
where $\phi_0$, $\theta_0$, and $\mathbf{u}_0$ are the homogeneous volume fraction, the homogeneous temperature, and the linear velocity field, respectively.
Here, we add small fluctuations, $\hat{\phi}$, $\hat{\theta}$, and $\hat{\mathbf{u}}=(\hat{u}_x,\hat{u}_y,\hat{u}_z)$,
to the homogeneous fields as $\phi=\phi_0+\hat{\phi}$, $\theta=\theta_0+\hat{\theta}$, and $\mathbf{u}=\mathbf{u}_0+\hat{\mathbf{u}}$, respectively.
In the following, we linearize the pressure, shear viscosity, and dissipation rate as
\begin{eqnarray}
p(\phi,\theta) &\simeq& p_0 + p_\phi\hat{\phi} + p_\theta\hat{\theta}~,\label{app:eq:expansion_p}\\
\eta(\phi,\theta) &\simeq& \eta_0+\eta_\phi\hat{\phi}+\eta_\theta\hat{\theta}~,\label{app:eq:expansion_eta}\\
\zeta(\phi,\theta) &\simeq& \zeta_0+\zeta_\phi\hat{\phi}+\zeta_\theta\hat{\theta}~,\label{app:eq:expansion_zeta}
\end{eqnarray}
respectively, where $p_0$, $\eta_0$, and $\zeta_0$ are the homogeneous values of $p$, $\eta$, and $\zeta$, respectively,
and the derivatives in the homogeneous state,\ i.e.\ $p_\phi$, $p_\theta$, $\eta_\phi$, $\eta_\theta$, $\zeta_\phi$, and $\zeta_\theta$, are listed in Table \ref{tab:derivatives}.
%
\begin{table}[h]\small
\caption{\
Dimensionless coefficients in Eqs.\ (\ref{app:eq:expansion_p})-(\ref{app:eq:expansion_zeta}).}
\label{tab:derivatives}
\begin{tabular*}{0.5\textwidth}{@{\extracolsep{\fill}}l}
\hline
$p_\phi = \theta_0/\left(1-\phi_0\right)^2-2\phi_0$~,\\
$p_\theta = \phi_0/(1-\phi_0)$~,\\
$\eta_\phi = (5\sqrt{\theta_0}/16\sqrt{\pi})\partial f_\eta(\phi_0)/\partial\phi_0$~,\\
$\eta_\theta = (5/32\sqrt{\pi\theta_0})f_\eta(\phi_0)$~,\\
$\zeta_\phi = \left\{(3h_1+32)/24\right\}\sqrt{\pi\theta_0}(1-e^2)(\chi+\phi_0\chi_\phi)$~,\\
$\zeta_\theta = \left\{(3h_1+32)/48\right\}\sqrt{\pi/\theta_0}(1-e^2)\phi_0\chi$~,\\
$\omega_\phi = s^2\eta_\phi-(d_\mathrm{m}/2)\left(\zeta_0+\phi_0\zeta_\phi\right)\theta_0$~,\\
$\omega_\theta = s^2\eta_\theta-(d_\mathrm{m}/2)\left(\zeta_0+\theta_0\zeta_\theta\right)\phi_0$~,\\
$\chi_\phi = -36\pi(\pi\phi-15)/(\pi\phi-6)^4$~,\\
$\nu_\phi = (\pi/5)(1+e)(\chi+\phi\chi_\phi)$~,\\
$\partial f_\eta/\partial\phi = (h_3-h_2)^{-1}\left(1+2\nu/3\right)\partial\chi^{-1}/\partial\phi$\\
$\hspace{1cm}+(2/3)f_\eta^\mathrm{k}(\phi)\nu_\phi+(3/45)(32-h_1)\left(\nu+\phi\nu_\phi\right)$~,\\
$\partial\chi^{-1}/\partial\phi = \pi(\pi\phi-15)(\pi\phi-6)^2/\left\{9(\pi\phi-12)^2\right\}$~.\\
\hline
\end{tabular*}
\end{table}
%
\subsection{Material derivatives and the continuity equation}
The material derivatives in Eqs.\ (\ref{eq:he1})-(\ref{eq:he3}) are linearized as
\begin{eqnarray}
\frac{\mathcal{D}X}{\mathcal{D}t} &\simeq& \frac{\partial\hat{X}}{\partial t} + sy\nabla_x\hat{X}\hspace{5mm} (X=\phi,\theta,u_y,u_z)~,\label{app:eq:lin_mat1}\\
\frac{\mathcal{D}u_x}{\mathcal{D}t} &\simeq& \frac{\partial\hat{u}_x}{\partial t} + sy\nabla_x\hat{u}_x + s\hat{u}_y~,\label{app:eq:lin_mat2}
\end{eqnarray}
where the last term on the right-hand-side of Eq.\ (\ref{app:eq:lin_mat2}) is the result of $\hat{\mathbf{u}}\cdot\nabla(sy)=s\hat{u}_y$.
Then, the left-hand-sides of Eqs.\ (\ref{eq:he1})-(\ref{eq:he3}) are linearized as
\begin{eqnarray}
\frac{\mathcal{D}\phi}{\mathcal{D}t} &\simeq& \frac{\partial\hat{\phi}}{\partial t} + sy\nabla_x\hat{\phi}~,\\
\phi\frac{\mathcal{D}u_i}{\mathcal{D}t} &\simeq& \phi_0\left(\frac{\partial\hat{u}_i}{\partial t} + sy\nabla_x\hat{u}_i + \delta_{ix}s\hat{u}_y\right)~,\\
\frac{d_\mathrm{m}}{2}\phi\frac{\mathcal{D}\theta}{\mathcal{D}t} &\simeq& \frac{d_\mathrm{m}}{2}\phi_0\left(\frac{\partial\hat{\theta}}{\partial t} + sy\nabla_x\hat{\theta}\right)~,
\end{eqnarray}
respectively.

Because the homogeneous solution is incompressive,\ i.e.\ $\nabla\cdot\mathbf{u}_0=0$,
the velocity gradient is linearized as $\nabla\cdot\mathbf{u}\simeq\nabla\cdot\hat{\mathbf{u}}$
so that the linearized continuity equation is given by Eq.\ (\ref{eq:lin_he1}).
\subsection{The equation of motion}
Next, we linearize the dimensionless equation of motion, Eq.\ (\ref{eq:he2}).
Since the diagonal part of dimensionless reversible stress tensor is linearized as $p_1 \simeq -2\phi_0\theta_0\nabla^2\hat{\phi}$,
the dimensionless reversible stress tensor is reduced to $\pi_{ij}\simeq p_0\delta_{ij}+\hat{\pi}_{ij}$ with the first order term,
\begin{equation}
\hat{\pi}_{ij} = \left\{\left(p_\phi-2\phi_0\theta_0\nabla^2\right)\hat{\phi}+p_\theta\hat{\theta}\right\}\delta_{ij}~,
\end{equation}
where its off-diagonal part is zero.
The diagonal part of dimensionless viscous stress tensor is linearized as $\tau_{ii}\simeq\hat{\tau}_{ii}$ with
\begin{equation}
\hat{\tau}_{ii}=2\eta_0\nabla_i\hat{u}_i + \left(\xi_0-\frac{2}{d_\mathrm{m}}\eta_0\right)\nabla_k\hat{u}_k~,
\end{equation}
where the dimensionless bulk viscosity, $\xi_0$, is defined in the homogeneous state
(note that the first term on the right-hand-side, $\nabla_i\hat{u}_i$, should not be summed over the subscript, $i$).
The dimensionless shear stress and the other off-diagonal parts of the dimensionless viscous stress are linearized as
$\tau_{xy}\simeq s\eta_0+\hat{\tau}_{xy}$ and $\tau_{ij}\simeq\hat{\tau}_{ij}$~($ij\neq xy,yx$), respectively,
where the first order terms are given by
\begin{eqnarray}
\hat{\tau}_{xy} &=& \eta_0\left(\nabla_x\hat{u}_y+\nabla_y\hat{u}_x\right)+s(\eta_\phi\hat{\phi}+\eta_\theta\hat{\theta})~,\\
\hat{\tau}_{ij} &=& \eta_0\left(\nabla_i\hat{u}_j+\nabla_j\hat{u}_i\right)\hspace{5mm} (ij\neq xy,yx)~,
\end{eqnarray}
respectively.
Then, the dimensionless stress gradient is linearized as
\begin{eqnarray}
\nabla_j\sigma_{xj}
&\simeq& \left\{s\eta_\phi\nabla_y-\left(p_\phi-2\phi_0\theta_0\nabla^2\right)\nabla_x\right\}\hat{\phi}\nonumber\\
& &+ \left(s\eta_\theta\nabla_y-p_\theta\nabla_x\right)\hat{\theta} + \left(\upsilon_0\nabla_x^2+\eta_0\nabla^2\right)\hat{u}_x\nonumber\\
& &+ \upsilon_0\nabla_x\nabla_y\hat{u}_y + \upsilon_0\nabla_z\nabla_x\hat{u}_z~,\nonumber
\end{eqnarray}
where we have introduced the sum of dimensionless viscosities as $\upsilon_0 = \left(1-2/d_\mathrm{m}\right)\eta_0 + \xi_0$.
Therefore, the linearized dimensionless equations of motion are found to be Eqs.\ (\ref{eq:lin_he2x})-(\ref{eq:lin_he2z}).
\subsection{The equation of temperature}
Finally, we linearize the equation of dimensionless temperature, Eq.\ (\ref{eq:he3}).
It is readily found that $\sigma_{ii}\nabla_i u_i \simeq -p_0\nabla_k\hat{u}_k$,
$\sigma_{ij}\nabla_i u_j \simeq 0$ ($i\neq j$ and $ij\neq xy,yx$), $\sigma_{xy}\nabla_xu_y\simeq s\eta_0\nabla_x\hat{u}_y$,
and $\sigma_{yx}\nabla_yu_x\simeq s^2\eta_0+s^2(\eta_\phi\hat{\phi}+\eta_\theta\hat{\theta})+2s\eta_0\nabla_y\hat{u}_x+s\eta_0\nabla_x\hat{u}_y$.
Thus, the dimensionless total power is linearized as
\begin{equation}
\sigma_{ij}\nabla_i u_j
\simeq s^2\eta_0 + s^2(\eta_\phi\hat{\phi}+\eta_\theta\hat{\theta})
+2s\eta_0\left(\nabla_y\hat{u}_x+\nabla_x\hat{u}_y\right)-p_0\nabla_k\hat{u}_k~.\nonumber
\end{equation}
The heat current and correction term for the dissipation rate are linearized as
$\nabla_iq_i\simeq\kappa_0\nabla^2\hat{\theta}+\mu_0\nabla^2\hat{\phi}$ and $-(1-e^2)f_\zeta(\phi)\nabla_ku_k\simeq-(1-e^2)f_\zeta(\phi_0)\nabla_k\hat{u}_k$, respectively,
where $\kappa_0$ and $\mu_0$ are the dimensionless thermal conductivity
and dimensionless transport coefficient proportional to the density gradient in the homogeneous state, respectively.
Thus, the linearized dimensionless equation of temperature is given by Eq.\ (\ref{eq:lin_he3}).

Note that the zero-th order equation, $\omega_0=s^2\eta_\phi-(d_\mathrm{m}/2)\phi_0\theta_0\zeta_0=0$,
represents the balance between the viscous heating and energy dissipation in the bulk,
where the dimensionless homogeneous temperature, $\theta_0$, is given by Eq.\ (\ref{eq:theta0}).
\subsection{The Fourier transforms}
\label{app:sub:Fourier}
We introduce the Fourier transforms of the small fluctuations as Eqs.\ (\ref{eq:Fourier_phi})-(\ref{eq:Fourier_u}).
In the linearized continuity equation (\ref{eq:lin_he1}), the second term on the left-hand-side explicitly depends on the $y$-coordinate, which is transformed as
\begin{equation}
sy\nabla_x\hat{\phi} = -\int sk_x\frac{\partial\phi_\mathbf{k}}{\partial k_y}e^{i\mathbf{k}\cdot\mathbf{r}}d\mathbf{k}~,
\end{equation}
where we have used $\int_{-\infty}^\infty\left(\partial\phi_\mathbf{k}e^{i\mathbf{k}\cdot\mathbf{r}}/\partial k_y\right)dk_y=0$.
Therefore, the linearized continuity equation in the Fourier space is given by \cite{onuki}
\begin{equation}
\left(\frac{\partial}{\partial t}-sk_x\frac{\partial}{\partial k_y}\right)\phi_\mathbf{k} = \phi_0\mathbf{k}\cdot\mathbf{u}_\mathbf{k}~.
\label{eq:fou_he1}
\end{equation}
Similarly, we transform the linearized equations (\ref{eq:lin_he2x})-(\ref{eq:lin_he3}) into the Fourier space to find
\begin{equation}
\left(\frac{\partial}{\partial t}-sk_x\frac{\partial}{\partial k_y}\right)\mathbf{\varphi}_\mathbf{k} = \mathcal{L}\mathbf{\varphi}_\mathbf{k}~,
\label{eq:lin_L}
\end{equation}
where we have introduced a vector of the Fourier coefficients as
$\mathbf{\varphi}_\mathbf{k}=\left(\phi_\mathbf{k},\theta_\mathbf{k},u_{x\mathbf{k}},u_{y\mathbf{k}},u_{z\mathbf{k}}\right)^\mathrm{T}$
and each component of a $5\times5$ matrix,
\begin{equation}
\mathcal{L}=\left(\mathcal{L}_{\alpha\beta}\right)~,
\label{eq:L}
\end{equation}
is defined as
\begin{eqnarray}
\mathcal{L}_{11} &=& \mathcal{L}_{12} = 0~,\hspace{2mm}
\mathcal{L}_{13} = \phi_0k_x~,\hspace{2mm}
\mathcal{L}_{14} = \phi_0k_y~,\hspace{2mm}
\mathcal{L}_{15} = \phi_0k_z~,\nonumber\\
\mathcal{L}_{21} &=& \bar{\omega}_\phi-\bar{\mu}_0 k^2~,\hspace{2mm}
\mathcal{L}_{22} = \bar{\omega}_\theta-\bar{\kappa}_0 k^2~,\hspace{2mm}
\mathcal{L}_{23} = ak_x - bsk_y~,\nonumber\\
\mathcal{L}_{24} &=& ak_y - bsk_x~,\hspace{2mm}
\mathcal{L}_{25} = ak_z~,\nonumber\\
\mathcal{L}_{31} &=& s\bar{\eta}_\phi k_y-\left(\bar{p}_\phi+2\theta_0k^2\right)k_x~,\hspace{2mm}
\mathcal{L}_{32} = s\bar{\eta}_\theta k_y - \bar{p}_\theta k_x~,\nonumber\\
\mathcal{L}_{33} &=& -\bar{\upsilon}_0k_x^2-\bar{\eta}_0k^2~,\hspace{2mm}
\mathcal{L}_{34} = -\bar{\upsilon}_0k_xk_y-s~,\hspace{2mm}
\mathcal{L}_{35} = -\bar{\upsilon}_0k_zk_x~,\nonumber\\
\mathcal{L}_{41} &=& s\bar{\eta}_\phi k_x-\left(\bar{p}_\phi+2\theta_0k^2\right)k_y~,\hspace{2mm}
\mathcal{L}_{42} = s\bar{\eta}_\theta k_x - \bar{p}_\theta k_y~,\nonumber\\
\mathcal{L}_{43} &=& -\bar{\upsilon}_0k_xk_y~,\hspace{2mm}
\mathcal{L}_{44} = -\bar{\upsilon}_0k_y^2-\bar{\eta}_0k^2~,\hspace{2mm}
\mathcal{L}_{45} = -\bar{\upsilon}_0k_yk_z~,\nonumber\\
\mathcal{L}_{51} &=& -\left(\bar{p}_\phi+2\theta_0k^2\right)k_z~,\hspace{2mm}
\mathcal{L}_{52} = -\bar{p}_\theta k_z~,\hspace{2mm}
\mathcal{L}_{53} = -\bar{\upsilon}_0k_zk_x~,\nonumber\\
\mathcal{L}_{54} &=& -\bar{\upsilon}_0k_yk_z~,\hspace{2mm}
\mathcal{L}_{55} = -\bar{\upsilon}_0k_z^2-\bar{\eta}_0k^2~,\nonumber
\end{eqnarray}
with the dimensionless wave number, $k=|\mathbf{k}|$.
%
\subsection{Transverse and longitudinal modes}
We represent the vector, $\mathbf{\varphi}_\mathbf{k}$, in a linear combination of the unit vectors,
\begin{eqnarray}
\mathbf{n}_1 &=& (1,0,0,0,0)^\mathrm{T}~,\\
\mathbf{n}_2 &=& (0,1,0,0,0)^\mathrm{T}~,\\
\mathbf{n}_3 &=& (0,0,e_x,e_y,e_z)^\mathrm{T}~,\\
\mathbf{n}_4 &=& (0,0,-e_x^\perp e_y, e_\perp, -e_ye_z^\perp)^\mathrm{T}~,\\
\mathbf{n}_5 &=& (0,0,-e_z^\perp, 0, e_x^\perp)^\mathrm{T}~,
\end{eqnarray}
as $\mathbf{\varphi}_\mathbf{k}=\sum_{j=1}^5 a_j\mathbf{n}_j$,
where the scaled wave numbers are introduced as $e_j\equiv k_j/k$, $e_j^\perp\equiv k_j/k_\perp$ $(j=x,y,z)$, and $e_\perp\equiv k_\perp/k$ with $k_\perp\equiv(k^2-k_y^2)^{1/2}$.
Here, $\mathbf{n}_3$ is parallel to the dimensionless wave number vector, while $\mathbf{n}_4$ and $\mathbf{n}_5$ are perpendicular to it,\ i.e.\
$\mathbf{n}_3\cdot\mathbf{k}=k$ and $\mathbf{n}_4\cdot\mathbf{k}=\mathbf{n}_5\cdot\mathbf{k}=0$.

A new vector, $\hat{\mathbf{\varphi}}_\mathbf{k}=(a_1,a_2,a_3,a_4,a_5)^\mathrm{T}$, is defined as
$\mathbf{\varphi}_\mathbf{k}=U\hat{\mathbf{\varphi}}_\mathbf{k}$ with a matrix, $U=\left(\mathbf{n}_1,\mathbf{n}_2,\mathbf{n}_3,\mathbf{n}_4,\mathbf{n}_5\right)$.
Then, the linearized hydrodynamic equation (\ref{eq:lin_L}) is transformed as
\footnote[7]{The scaled wave numbers  satisfy the relations, $e_x^2+e_y^2+e_z^2=1$, $e_x^2+e_z^2=e_\perp^2$, $e_xe_x^\perp+e_ze_z^\perp=e_\perp$, and $e_\perp e_j^\perp=e_j$.
Because the unit vectors, $\mathbf{n}_j$~$(j=1,\dots,5)$, are orthonormal, the inverse matrix of $U$ is equal to the transposed one,\ i.e.\ $U^{-1}=U^T$.}
\begin{equation}
U^{-1}\left(\frac{\partial}{\partial t}-sk_x\frac{\partial}{\partial k_y}\right)\left(U\hat{\mathbf{\varphi}}_\mathbf{k}\right)
= \left(U^{-1}\mathcal{L}U\right)\hat{\mathbf{\varphi}}_\mathbf{k}~,
\end{equation}
where the left-hand-side is reduced to
\begin{equation}
\mathrm{(l.h.s)}
= \frac{\partial\hat{\mathbf{\varphi}}_\mathbf{k}}{\partial t}-sk_x\frac{\partial\hat{\mathbf{\varphi}}_\mathbf{k}}{\partial k_y}
-\left[sk_xU^{-1}\frac{\partial U}{\partial k_y}\right]\hat{\mathbf{\varphi}}_\mathbf{k}
\end{equation}
with a matrix,
\begin{equation}
sk_xU^{-1}\frac{\partial U}{\partial k_y} =
\begin{pmatrix}
	0 & 0 & 0 & 0 & 0\\
	0 & 0 & 0 & 0 & 0\\
	0 & 0 & 0 &-se_xe_\perp & 0\\
	0 & 0 & se_xe_\perp & 0 & 0\\
	0 & 0 & 0 & 0 & 0
\end{pmatrix}~.
\end{equation}
Therefore, the linearized hydrodynamic equations are rewritten as
\begin{equation}
\left(\frac{\partial}{\partial t}-sk_x\frac{\partial}{\partial k_y}\right)\hat{\mathbf{\varphi}}_\mathbf{k}
= P\hat{\mathbf{\varphi}}_\mathbf{k}~,
\end{equation}
where each component of the matrix, $P=U^{-1}\mathcal{L}U+sk_xU^{-1}(\partial U/\partial k_y)\equiv\left(P_{\alpha\beta}\right)$, is given by
\begin{eqnarray}
P_{11} &=& P_{12} = P_{14} = P_{15} = 0~,\hspace{2mm} P_{13} = \phi_0k~,\nonumber\\
P_{21} &=& \bar{\omega}_\phi-\bar{\mu}_0 k^2~,\hspace{2mm}
P_{22} = \bar{\omega}_\theta-\bar{\kappa}_0 k^2~,\nonumber\\
P_{23} &=& ak - 2bse_xk_y~,\hspace{2mm}
P_{24} = bsk_x(e_ye_y^\perp-e_\perp)~,\hspace{2mm}
P_{25} = bsk_ye_z^\perp~,\nonumber\\
P_{31} &=& 2s\bar{\eta}_\phi e_xk_y-\left(\bar{p}_\phi+2\theta_0k^2\right)k~,\hspace{2mm}
P_{32} = 2s\bar{\eta}_\theta e_xk_y-\bar{p}_\theta k~,\nonumber\\
P_{33} &=& -(\bar{\upsilon}_0+\bar{\eta}_0)k^2-se_xe_y~,\hspace{2mm}
P_{34} = -2se_xe_\perp~,\hspace{2mm}
P_{35} = 0~,\nonumber\\
P_{41} &=& s\bar{\eta}_\phi k_x(e_\perp-e_ye_y^\perp)~,\hspace{2mm}
P_{42} = s\bar{\eta}_\theta k_x(e_\perp-e_ye_y^\perp)~,\nonumber\\
P_{43} &=& se_x^\perp~,\hspace{2mm}
P_{44} = se_xe_y-\bar{\eta}_0k^2~,\hspace{2mm}
P_{45} = 0~,\nonumber\\
P_{51} &=& -s\bar{\eta}_\phi k_ye_z^\perp~,\hspace{2mm}
P_{52} = -s\bar{\eta}_\theta k_ye_z^\perp~,\nonumber\\
P_{53} &=& se_ye_z^\perp~,\hspace{2mm}
P_{54} = se_z~,\hspace{2mm}
P_{55} = -\bar{\eta}_0k^2~.
\end{eqnarray}
%
\section{Perturbation theory}
\label{app:perturbation}
In this Appendix, we perturbatively solve the eigenvalue problem,
\begin{equation}
\left(P+sk_x\frac{\partial}{\partial k_y}\right)\hat{\mathbf{\varphi}}_\mathbf{k} = \lambda\hat{\mathbf{\varphi}}_\mathbf{k}~,
\end{equation}
where we have introduced the growth rate which is equivalent to the eigenvalue, $\lambda$, as $\hat{\mathbf{\varphi}}_\mathbf{k}(t)\propto e^{\lambda t}$.
Here, we also define the left-eigenvector, $\hat{\mathbf{\psi}}_\mathbf{k}$, as
\begin{equation}
\left(P^\mathrm{T}+sk_x\frac{\partial}{\partial k_y}\right)\hat{\mathbf{\psi}}_\mathbf{k}^\mathrm{T} = \lambda\hat{\mathbf{\psi}}_\mathbf{k}^\mathrm{T}~.
\end{equation}

At first, we introduce a small parameter, $\epsilon$, for the perturbative calculations,
where the dimensionless wave numbers are scaled as
\begin{equation}
k=\epsilon q~,\hspace{1mm}k_\perp=\epsilon q_\perp~,\hspace{1mm}
k_x=\epsilon q_x~,\hspace{1mm}k_y=\epsilon q_y~,\hspace{1mm}k_z=\epsilon q_z~,
\end{equation}
which do not change the scaled wave numbers,\ i.e.\ $e_j=k_j/k=q_j/q$, $e_j^\perp=k_j/k_\perp=q_j/q_\perp$ $(j=x,y,z)$, and $e_\perp=k_\perp/k=q_\perp/q$.
The dimensionless shear rate and inelasticity are scaled as
\begin{equation}
s=\epsilon^2\bar{s}~,\hspace{4mm}1-e^2=\epsilon^4\varsigma~,
\end{equation}
respectively, so that the homogeneous temperature, $\theta_0\sim s^2/(1-e^2)=\bar{s}^2/\varsigma$, remains as finite.

Then, we expand the eigenvalue, right- and left-eigenvectors into the powers of $\epsilon$ as
\begin{eqnarray}
\lambda &=& \epsilon\lambda_1+\epsilon^2\lambda_2+\dots~,\\
\hat{\mathbf{\varphi}}_\mathbf{k} &=& \hat{\mathbf{\varphi}}_0+\epsilon\hat{\mathbf{\varphi}}_1+\epsilon^2\hat{\mathbf{\varphi}}_2+\dots~,\\
\hat{\mathbf{\psi}}_\mathbf{k} &=& \hat{\mathbf{\psi}}_0+\epsilon\hat{\mathbf{\psi}}_1+\epsilon^2\hat{\mathbf{\psi}}_2+\dots~,
\end{eqnarray}
respectively.
The matrix, $P$, is also expanded into the powers of $\epsilon$ as
\begin{eqnarray}
P &=& \epsilon qA+\epsilon^2\{q^2B+\bar{s}C\}+\epsilon^3\{q^3F+\bar{s}qG\}+\dots
\end{eqnarray}
with the matrices,
\begin{eqnarray}
A&=&
\begin{pmatrix}
0 & 0 & \phi_0 & 0 & 0 \\
0 & 0 & a_0 & 0 & 0 \\
-\bar{p}_\phi & -\bar{p}_\theta & 0 & 0 & 0 \\
0 & 0 & 0 & 0 & 0\\
0 & 0 & 0 & 0 & 0
\end{pmatrix}~,\nonumber\\
B&=&
\begin{pmatrix}
0 & 0 & 0 & 0 & 0 \\
0 & -\bar{\kappa}_0 & 0 & 0 & 0 \\
0 & 0 & -(\bar{\upsilon}_0+\bar{\eta}_0) & 0 & 0 \\
0 & 0 & 0 & -\bar{\eta}_0 & 0 \\
0 & 0 & 0 & 0 & -\bar{\eta}_0
\end{pmatrix}~,\nonumber\\
C&=&
\begin{pmatrix}
0 & 0 & 0 & 0 & 0 \\
0 & 0 & 0 & 0 & 0 \\
0 & 0 & -e_xe_y & -2e_xe_\perp & 0 \\
0 & 0 & e_x^\perp & e_xe_y & 0 \\
0 & 0 & e_ye_z^\perp & e_z & 0
\end{pmatrix}~.\nonumber
\end{eqnarray}
%
%
\subsection{The 1st order equation}
\label{app:sub:1st_order}
The first order equation is found to be
\begin{equation}
qA\hat{\varphi}_0 = \lambda_1\hat{\varphi}_0~,\hspace{5mm}
qA^\mathrm{T}\hat{\psi}_0^\mathrm{T} = \lambda_1\hat{\psi}_0^\mathrm{T}~,
\end{equation}
where the five eigenvalues are given by
\begin{equation}
\lambda_1^{(1)}=-ifq,\hspace{2mm}\lambda_1^{(2)}=ifq,\hspace{2mm}\lambda_1^{(3)}=\lambda_1^{(4)}=\lambda_1^{(5)}=0
\end{equation}
(note that the superscripts represent different eigenmodes).
Here, we have introduced a constant as
\begin{eqnarray}
f=\sqrt{a_0\bar{p}_\theta+\phi_0\bar{p}_\phi}=\sqrt{p_\phi+\frac{2p_\theta^2\theta_0}{d_\mathrm{m}\phi_0^2}}~.
\label{app:eq:f}
\end{eqnarray}
The corresponding right- and left-eigenvectors are found to be
\begin{eqnarray}
\varphi_0^{(1)} &=& \frac{1}{\sqrt{2}f}(\phi_0,a_0,-if,0,0)^\mathrm{T}~,\\
\varphi_0^{(2)} &=& \frac{1}{\sqrt{2}f}(\phi_0,a_0,+if,0,0)^\mathrm{T}~,\\
\varphi_0^{(3)} &=& \frac{1}{f}(\bar{p}_\theta,-\bar{p}_\phi,0,0,0)^\mathrm{T}~,\\
\varphi_0^{(4)} &=& (0,0,0,1,0)^\mathrm{T}~,\\
\varphi_0^{(5)} &=& (0,0,0,0,1)^\mathrm{T}~,
\end{eqnarray}
and
\begin{eqnarray}
\psi_0^{(1)} &=& \frac{1}{\sqrt{2}f}(\bar{p}_\phi,\bar{p}_\theta,+if,0,0)~,\\
\psi_0^{(2)} &=& \frac{1}{\sqrt{2}f}(\bar{p}_\phi,\bar{p}_\theta,-if,0,0)~,\\
\psi_0^{(3)} &=& \frac{1}{f}(a_0,-\phi_0,0,0,0)~,\\
\psi_0^{(4)} &=& (0,0,0,1,0)~,\\
\psi_0^{(5)} &=& (0,0,0,0,1)~,
\end{eqnarray}
respectively.
%
\subsection{The 2nd order equation}
\label{app:sub:2nd_order}
Because the three eigenvalues, $\lambda_1^{(l)}$~$(l=3,4,5)$, are degenerated to zero,
we replace the right-eigenvectors, $\varphi_0^{(l)}$, with a linear series \cite{saitoh1},
\begin{equation}
h_m^{(l)}\varphi_0^{(m)} \equiv h_3^{(l)}\varphi_0^{(3)}+h_4^{(l)}\varphi_0^{(4)}+h_5^{(l)}\varphi_0^{(5)}~,
\end{equation}
where the coefficients, $h_m^{(l)}$~$(m=3,4,5)$, will be determined in the following.
Then, the second order equation is found to be
\begin{equation}
qA\varphi_1^{(l)}+\left(\mathcal{M}+\bar{s}q_x\frac{\partial}{\partial q_y}\right)h_m^{(l)}\varphi_0^{(m)}=\lambda_2^{(l)}h_m^{(l)}\varphi_0^{(m)}~,
\label{app:eq:2nd-eq}
\end{equation}
where we have introduced $\mathcal{M}\equiv q^2B+\bar{s}C$.
Multiplying both sides by the left-eigenvectors, $\psi_0^{(n)}$~$(n=3,4,5)$,
we find that the first term on the left-hand-side vanishes (because of $\psi_0^{(n)}A=0$) and the second term on the left-hand-side is reduced to
\begin{eqnarray}
& & \psi_0^{(n)}\left(\mathcal{M}+\bar{s}q_x\frac{\partial}{\partial q_y}\right)h_m^{(l)}\varphi_0^{(m)}\nonumber\\
&=& \psi_0^{(n)}\mathcal{M}\varphi_0^{(m)}h_m^{(l)} + \bar{s}q_x\psi_0^{(n)}\varphi_0^{(m)}\frac{\partial h_m^{(l)}}{\partial q_y}\nonumber\\
&=& \psi_0^{(n)}\mathcal{M}\varphi_0^{(m)}h_m^{(l)} + \bar{s}q_x\frac{\partial h_n^{(l)}}{\partial q_y}~,
\end{eqnarray}
because $\varphi_0^{(m)}$ is independent of the wave number and we have used $\psi_0^{(n)}\varphi_0^{(m)}=\delta_{nm}$.
Thus, Eq.\ (\ref{app:eq:2nd-eq}) becomes a \emph{linear simultaneous differential equation} for the coefficients, $h_n^{(l)}$,\ i.e.\
\begin{equation}
\bar{s}q_x\frac{\partial h_n^{(l)}}{\partial q_y} + \psi_0^{(n)}\mathcal{M}\varphi_0^{(m)}h_m^{(l)}= \lambda_2^{(l)}h_n^{(l)}~,
\label{app:eq:simultaneous}
\end{equation}
where the matrix is explicitly given by
\begin{eqnarray}
\psi_0^{(n)}\mathcal{M}\varphi_0^{(m)} &=&
\begin{pmatrix}
-gq^2 & 0 & 0 \\
0 & \bar{s}e_xe_y-\bar{\eta}_0q^2 & 0 \\
0 & \bar{s}e_z & -\bar{\eta}_0q^2
\end{pmatrix}
\end{eqnarray}
with a constant,
\begin{equation}
g = \frac{2\kappa_0p_\phi}{d_\mathrm{m}\phi_0f^2}~.
\label{app:eq:g}
\end{equation}

The differential equation (\ref{app:eq:simultaneous}) for $n=3$ is given by
\begin{equation}
\bar{s}q_x\frac{\partial}{\partial q_y}h_3^{(l)} = \left(\lambda_2^{(l)}+gq^2\right)h_3^{(l)}~,
\label{app:eq:diffeq_n3}
\end{equation}
where the eigenvalue and solutions for $l=3$ are readily found to be
\begin{equation}
\lambda_2^{(3)} = -gq^2,\hspace{5mm} h_3^{(3)}=1,\hspace{5mm} h_4^{(3)}=h_5^{(3)}=0~,
\end{equation}
respectively, which also satisfy the differential equations (\ref{app:eq:simultaneous}) for $n=4,5$.
The other eigenvalues, $\lambda_2^{(4)}$ and $\lambda_2^{(5)}$, are different from $\lambda_2^{(3)}$
so that $h_3^{(4)}=h_3^{(5)}=0$ to satisfy the first differential equation (\ref{app:eq:diffeq_n3}).
Then, the differential equations (\ref{app:eq:simultaneous}) for $n=4,5$ are given by
\begin{eqnarray}
\bar{s}q_x\frac{\partial}{\partial q_y}h_4^{(l)} &=& \left(\lambda_2^{(l)}-\bar{s}e_xe_y+\bar{\eta}_0q^2\right)h_4^{(l)}~,\label{app:eq:diffeq_n4}\\
\bar{s}q_x\frac{\partial}{\partial q_y}h_5^{(l)} &=& \left(\lambda_2^{(l)}+\bar{\eta}_0q^2\right)h_5^{(l)}-\bar{s}e_z h_4^{(l)}~,\label{app:eq:diffeq_n5}
\end{eqnarray}
respectively ($l=4,5$).
Here, we choose the eigenvalue for $l=4$ as $\lambda_2^{(4)} = \bar{s}e_xe_y-\bar{\eta}_0q^2$ to find the solution as $h_4^{(4)}=1$.
Then, the last differential equation (\ref{app:eq:diffeq_n5}) is reduced to
\begin{eqnarray}
\frac{\partial}{\partial q_y}h_5^{(4)} = \frac{q_y}{q^2}h_5^{(4)}-\frac{q_z}{qq_x}~,
\end{eqnarray}
where its solution is found to be
\begin{equation}
h_5^{(4)} = \frac{qq_z}{q_xq_\perp}\tan^{-1}\left(-\frac{q_y}{q_\perp}\right)~.
\end{equation}
Similarly, if we choose the eigenvalue, $\lambda_2^{(5)} = -\bar{\eta}_0q^2$, the differential equations (\ref{app:eq:diffeq_n4}) and (\ref{app:eq:diffeq_n5}) are reduced to
\begin{equation}
\frac{\partial}{\partial q_y}h_4^{(5)} = -\frac{q_y}{q^2}h_4^{(5)}~,\hspace{5mm}
\frac{\partial}{\partial q_y}h_5^{(5)} = -\frac{q_z}{qq_x}h_4^{(5)}~,
\end{equation}
respectively, where their solutions are readily found to be $h_4^{(5)}=0$ and $h_5^{(5)}=1$.
\subsection{Unstable mode}
\label{app:sub:unstable_mode}
In summary, the eigenvalues and corresponding right-eigenvectors are given by
\begin{eqnarray}
\lambda^{(1)} &=& -\lambda^{(2)} = -ifk~,\\
\lambda^{(3)} &=& -gk^2~,\label{app:eq:isotropic_eigenvalue}\\
\lambda^{(4)} &=& se_xe_y-\bar{\eta}_0k^2~,\label{app:eq:anisotropic_eigenvalue}\\
\lambda^{(5)} &=& -\bar{\eta}_0k^2~,
\end{eqnarray}
and
\begin{eqnarray}
\varphi_0^{(1)} &=& \frac{1}{\sqrt{2}f}(\phi_0,a_0,-if,0,0)^\mathrm{T}~,\\
\varphi_0^{(2)} &=& \frac{1}{\sqrt{2}f}(\phi_0,a_0,+if,0,0)^\mathrm{T}~,\\
\varphi_0^{(3)} &=& \frac{1}{f}(\bar{p}_\theta,-\bar{p}_\phi,0,0,0)^\mathrm{T}~,\\
\varphi_0^{(4)} &+& \frac{kk_z}{k_xk_\perp}\tan^{-1}\left(-\frac{k_y}{k_\perp}\right)\varphi_0^{(5)}\nonumber\\
&=& \left(0,0,0,1,\frac{kk_z}{k_xk_\perp}\tan^{-1}\left(-\frac{k_y}{k_\perp}\right)\right)^\mathrm{T}~,\label{app:eq:anisotropic_eigenvector}\\
\varphi_0^{(5)} &=& (0,0,0,0,1)^\mathrm{T}~,
\end{eqnarray}
respectively, where these results are consistent with those in Ref.\ \cite{LutskoDufty}.

Because the first two eigenvalues, $\lambda^{(1)}$ and $\lambda^{(2)}$, represent propagating modes and the fifth eigenvalue is negative, $\lambda^{(5)}<0$,
the third and fourth eigenvalues, $\lambda^{(3)}$ and $\lambda^{(4)}$, can be unstable,
where $\lambda^{(3)}$ and $\lambda^{(4)}$ are \emph{isotropic} and \emph{anisotropic} in the Fourier space, respectively.
The isotropic eigenvalue, $\lambda^{(3)}$, is positive if $g<0$,\ i.e.\
the system is \emph{thermodynamically unstable}, $p_\phi<0$, which could happen in the van der Waals model.
On the other hand, the anisotropic eigenvalue, $\lambda^{(4)}$, is positive if $se_xe_y>\bar{\eta}_0k^2$,\ i.e.\
\begin{equation}
\frac{k_xk_y}{k^4}>\left\{\frac{5\sqrt{15}f_\eta(\phi_0)^{3/2}}{16\pi\phi_0^2\sqrt{d_\mathrm{m}(3h_1+32)\chi(\phi_0)}}\right\}\frac{1}{\sqrt{1-e^2}}~.
\label{app:eq:anisotropic_unstable_condition}
\end{equation}
Eq.\ (\ref{app:eq:anisotropic_unstable_condition}) corresponds to the \emph{shear-induced instability} for usual (dry) granular shear flows \cite{saitoh1},
where we have used $\bar{\eta}_0=\eta_0/\phi_0$ and Eq.\ (\ref{eq:theta0}).
In cohesive granular shear flows, the thermodynamic instability, $p_\phi<0$, and shear-induced instability, Eq.\ (\ref{app:eq:anisotropic_unstable_condition}),
compete with each other, where the second criterion, Eq.\ (\ref{app:eq:anisotropic_unstable_condition}),
also depends on the system size, $L$, through the wave numbers, $k_x,k_y,k\sim1/L$.
Such a system size dependence of the shear-induced instability is consistent with the previous stability analyses of dry granular shear flows \cite{linear5,linear6}.
Note that the isotropic eigenvalue is always larger than the anisotropic one,\ i.e.\ $\lambda^{(3)}>\lambda^{(4)}$,
in our system size, $L=50d$, with the range of control parameters used in this study.
%
%
%
\bibliography{multi}
\bibliographystyle{rsc}
\end{document}